\documentclass[journal ]{new-aiaa}
\usepackage[utf8]{inputenc}
\usepackage{textcomp}

\usepackage{graphicx}
\usepackage{amsmath}

\usepackage{amsthm}
\theoremstyle{remark}
\newtheorem{remark}{Remark}
\usepackage[version=4]{mhchem}
\usepackage{siunitx}
\usepackage{longtable,tabularx}
\usepackage[percent]{overpic}
\usepackage{pict2e}
\usepackage{subcaption}

\usepackage{algorithm}
\usepackage{algpseudocode}
\usepackage{mathtools}
\usepackage{cleveref}[2012/02/15]
\allowdisplaybreaks

\usepackage{bm}

\usepackage{comment}

\newcommand{\norm}[1]{\left\lVert#1\right\rVert}

\usepackage{xcolor}

\makeatletter
\newcommand\footnoteref[1]{\protected@xdef\@thefnmark{\ref{#1}}\@footnotemark}
\makeatother

\newcommand{\filippos}[1]{\textcolor{red}{[Filippos: #1]}}

\makeatletter
\@ifundefined{thefootnote}{%
  \newcounter{footnote}%
  \renewcommand\thefootnote{\arabic{footnote}}%
}{}
\makeatother

\newcommand{\mf}[1]{\mathbf{#1}}
\newcommand{\mr}[1]{\mathrm{#1}}

\title{Cislunar Pursuit–Evasion Game on Periodic and Quasi-Periodic Orbits}

\author{Quentin Rommel$^{\star}$\footnote{Graduate Research Assistant, Department of Aerospace Engineering \& Engineering Mechanics, e-mail: quentin.rommel@utexas.edu}, Filippos Fotiadis$^{\star}$\footnote{Postdoctoral Researcher, Oden Institute for Computational Engineering \& Sciences, e-mail: ffotiadis@utexas.edu}, Cade Armstrong \footnote{Graduate Research Assistant, Department of Aerospace Engineering \& Engineering Mechanics, e-mail: cadearmstrong@utexas.edu}, Luke Peterson\footnote{Assistant Professor, Department of Aerospace Engineering \& Engineering Mechanics, e-mail: ltp@utexas.edu}, Ufuk Topcu\footnote{Professor, Department of Aerospace Engineering \& Engineering Mechanics, e-mail: utopcu@utexas.edu}}
\affil{The University of Texas at Austin, Austin, TX, 78712}
\begin{document}

\maketitle

\makeatletter
\let\svthefootnote\thefootnote
\let\thefootnote\relax
\footnotetext{$^{\star}$ Equal contribution}
\let\thefootnote\svthefootnote
\makeatother

\begin{abstract}
Cislunar spacecraft operate in nonlinear and unstable environments that make defensive maneuver planning difficult. We formulate cislunar spacecraft pursuit and evasion as a zero-sum differential game in the circular restricted three-body problem. Each spacecraft controls its thrust and reference orbit phase, enabling motion along periodic orbits and across quasi-periodic tori while remaining near the reference. We solve the game using a constrained discrete-time differential dynamic programming method that enforces hard input constraints. We propose a shared time regularization to synchronize both spacecraft while refining the discretization near close lunar passages. Numerical results on periodic and quasi-periodic orbits show that phase control improves maneuvering flexibility while limiting departure from the reference orbit. The discrete-time method also provides a large computational advantage over a continuous-time formulation. Comparisons between quasi-halo and quasi-near-rectilinear halo orbits show that close lunar passages create larger escape opportunities but also increase the sensitivity of the encounter. These results show that reference orbit geometry is an important part of defensive cislunar mission design.
\end{abstract}
\section*{Nomenclature}
{\renewcommand\arraystretch{1.0}
\noindent\begin{longtable*}{@{}l @{\quad=\quad} l@{}}
CR3BP & Circular Restricted Three-Body Problem \\
CT-DDP & Continuous-Time Differential Dynamic Programming \\
DDP & Differential Dynamic Programming \\
DT-DDP & Discrete-Time Differential Dynamic Programming \\
GT-DDP & Game-Theoretic Differential Dynamic Programming \\
HJI & Hamilton--Jacobi--Isaacs \\
KS & Kustaanheimo--Stiefel \\
$L_1/L_2$ & Earth--Moon collinear Lagrange points \\
LU & Length Unit \\
NRHO & Near-Rectilinear Halo Orbit \\
PO & Periodic Orbit \\
QPO & Quasi-Periodic Orbit \\
TU & Time Unit \\
\end{longtable*}}
\setcounter{table}{0}

\section{Introduction}
\lettrine[lines=2,lhang=0.1,nindent=0em]{C}islunar space is becoming an important region for upcoming lunar and interplanetary missions~\cite{crusan2019nasa,baker2024comprehensive,holzinger2021primer}. Its remote and chaotic nature, however, creates new security challenges. Unlike low-Earth and geosynchronous missions, cislunar missions will be sparse and will depend on long-distance communication links~\cite{badura2023optimizing,kurtsecurity}. These conditions create potential single points of failure and give adversaries opportunities to pursue spacecraft, disrupt communications, or exploit unstable orbital dynamics to drive a target away from its nominal orbit. A safe and sustained presence in cislunar space therefore requires defensive strategies that allow spacecraft to respond to hostile assets.

Recent in-orbit maneuvers show the need for such strategies. In 2015, for example, Russia's ``Luch/Olymp-K'' satellite maneuvered unusually close to commercial communication satellites in geostationary orbit~\cite{roberts2024method,sankaran2022russia,roberts2020sustainable} illustrating the risks posed by uncooperative proximity operations. Extending these concerns to cislunar space, a spacecraft may face three broad pursuit threats.
The first is proximity-based interference, in which an attacker approaches the target to jam communications or degrade onboard sensing. Because ground-station links are weak at lunar distances, even low-power interference can reduce the accuracy of communication-based orbit determination and increase the risk of mission loss. The second threat is direct interception, in which the attacker matches the target's position and often its velocity with the intent to collide. Between these cases lies rendezvous and inspection, in which an adversary remains near the target to observe it or apply small disturbances. 

Defensive maneuvering in cislunar space is constrained by more than thrust capability. Spacecraft are often required to remain near prescribed trajectories because they provide favorable communication, observation, illumination, and station-keeping properties. A large departure from the reference orbit may avoid an immediate threat but can compromise the broader mission. The relevant problem is therefore not olny whether a spacecraft can increase its distance from a pursuer, but whether it can avoid the encounter while preserving its intended orbital geometry. Adjusting the phase of the reference motion provides such a capability. On a periodic orbit, phase adjustment changes the spacecraft’s along-track location and the timing of future close approaches. Quasi-periodic tori extend the maneuvering capabilities to higher dimensions. The reference orbit is not only as a trajectory to track, its phase and geometry can also serve as defensive resources that allow a spacecraft to reshape an adversarial encounter while preserving the broader mission. 

Pursuit-evasion games provide a mathematical method for studying such adversarial encounters. They model the interaction as a zero-sum game in which the pursuer seeks interception while the evader seeks escape~\cite{bacsar1998dynamic,isaacs1999differential}. Prior work has studied spacecraft pursuit-evasion in Earth orbit under Keplerian dynamics~\cite{fu2025analytical,shen2018revisit,mehlman2024cat}. These methods do not extend directly to cislunar space, where spacecraft motion is commonly modeled by the circular restricted three-body problem (CR3BP)~\cite{koon2000dynamical,connor1984three,szebehely1967}. The CR3BP is nonlinear and unstable, which limits the use of classical two-body formulations, particularly those based on local orbital frames. Moreover, the nonlinear dynamics of the CR3BP make the resulting differential game difficult to solve analytically, motivating the use of numerical optimal control methods.

Differential dynamic programming (DDP) offers a suitable approach for trajectory optimization under nonlinear dynamics. Early work combined DDP with a Sundman transformation to solve long-duration, many-revolution low-thrust transfers by redistributing trajectory nodes along the orbit~\cite{aziz2018low}. Hybrid DDP formulations later addressed path constraints, variable flight time, moving targets, and transfers between libration-point orbits in the CR3BP~\cite{aziz2019hybrid}. More recent methods have improved computational speed using higher-order state transition tensors~\cite{boone2025rapid} and combined DDP with Pontryagin's minimum principle for constrained low-thrust transfers in cislunar space~\cite{sidhoum2026pontryagin}. These studies show that DDP can handle nonlinear and constrained spacecraft trajectory problems, but they focus on single-spacecraft optimal control rather than adversarial multi-agent games. Moreover, while several studies extend DDP to the multi-agent case (e.g., \cite{sun2018min}), they do not address pursuit-evasion under the nonlinear and unstable dynamics of the CR3BP.

Close lunar passages introduce an additional numerical challenge because the dynamics vary rapidly near the Moon. Time regularization has been used to redistribute integration nodes and improve numerical resolution in such regions~\cite{aziz2018low}. Leith et al. extended this idea to spacecraft trajectories under multi-body gravity, where the time transformation adapts the discretization across different gravitational bodies and reduces the risk of missing close flybys~\cite{leith2023time}. Related global regularization methods use a shared fictitious time for interacting multi-body systems~\cite{heggie1974global}. These methods, however, are designed for single-spacecraft trajectory optimization or general interacting-body simulations, and do not address the synchronization requirements of adversarial multi-agent trajectory optimization.

To address these challenges, we develop a framework for constrained cislunar pursuit-evasion in the CR3BP. The framework contains three main components. First, we augment the spacecraft controls with reference phase rates that allow the agents to change their locations along periodic orbits and, for quasi-periodic references, across two-dimensional invariant tori without requiring large departures from the reference geometry. Second, as an algorithmic contribution, we develop a constrained discrete-time game-theoretic DDP method that computes local saddle-point policies while enforcing hard thrust and phase-rate bounds. During the backward pass, the method extends active-set DDP \cite{tassa2014control} to a two-player saddle-point problem by jointly solving the free controls of both agents and applying the appropriate bound-release conditions for the minimizer and maximizer. Third, we introduce a shared multi-agent time regularization that preserves synchronization in physical time and automatically increases the joint temporal resolution when either spacecraft approaches the Moon. Together, these components make it possible to compute practical pursuit-evasion strategies that exploit cislunar orbit geometry while respecting control and tracking limits.

Numerical studies demonstrate both the capabilities of the proposed framework and the impact of its individual components. First, they show that controlled phasing can increase separation while keeping the spacecraft close to its reference orbit, demonstrating that the added phase variables provide a practical alternative to large thrust-driven departures. Second, they show that quasi-periodic references provide an additional defensive degree of freedom: the transverse torus phase can modify the local encounter geometry and create escape opportunities that are unavailable on a single periodic curve. Third, comparisons between quasi-halo and quasi-NRHO families show that close lunar passages can amplify evasive opportunities, but also make the outcome more sensitive to maneuver timing. The studies also demonstrate the computational benefit of the proposed method: the shared regularization maintains convergence near small perilunes, and the discrete-time implementation is 30.2 times faster than the continuous-time implementation in the periodic NRHO case.

A preliminary version of this work appeared in~\cite{fotiadis2026adversarial}. Relative to that version, the present paper extends the framework by moving from unconstrained continuous DDP to constrained discrete DDP, incorporating quasi-periodic reference orbits with two-dimensional phase control, introducing a synchronization-preserving multi-agent time regularization, and substantially expanding the numerical study.

\textbf{Structure.} The paper is organized as follows. We first introduce the CR3BP dynamics and the periodic and quasi-periodic reference orbits used in the pursuit-evasion scenarios, including their phase parameterizations and the construction of the quasi-periodic tori. We then formulate the zero-sum differential game, where each spacecraft controls its thrust and reference phase to perform along-track maneuvers on periodic orbits and move across invariant tori on quasi-periodic orbits. Next, we develop the shared multi-agent time regularization and the regularized equations of motion used to resolve close lunar approaches. We then present the constrained discrete-time game-theoretic DDP method, including the local saddle-point solution and active-set treatment of the control bounds. Finally, we evaluate the method on periodic NRHO and quasi-periodic orbit families, compare it with the continuous-time formulation, quantify the benefit of the second phase control, and study how lunar-approach geometry and orbital instability affect the pursuit-evasion outcome.

\textbf{Notation.}  
Let $\mathbb{T}$ denote the torus. The identity and zero matrices are denoted by $I_n$ and $0_n$, respectively. The symbols $\norm{x}$ and $\norm{x}_Q$ denote the $\ell_2$ norm and $Q$-weighted $\ell_2$ norm, respectively. Subscripts of $V,\phi,\bar{V},\bar{L},\bar{\Phi}$ denote partial derivatives with respect to the indicated variable, e.g.,
$V_\mf{x}=\nabla_\mf{x}V$, $V_{\mf{x}\mf{x}}=\nabla_\mf{x}^2V$, and $\bar{L}_\mf{w}=\nabla_\mf{w}\bar{L}$.

\section{Pursuit–Evasion in Cislunar Space Setup}
\subsection{Spacecraft Dynamics in Cislunar Space}\label{sec:CR3BP}
In cislunar space, the dynamics of a spacecraft can be described through the Circular Restricted Three-Body Problem (CR3BP). In this model, the spacecraft is assumed to be a point of negligible mass under the gravitational influence of two large bodies, the Earth and the Moon, that revolve in circular orbits around their common barycenter. 

To describe the CR3BP, let us denote the mass ratio $\mu=\frac{m_M}{m_E+m_M}$ and subsequently normalize the Earth-Moon system's masses and length, so that $m_E=1-\mu$ is the normalized Earth's mass, $m_M=\mu$ is the normalized Moon's mass, and so that the distance between the Earth and the Moon is equal to $1$. Consider also the rotating barycentric reference frame, so that the position of the Earth is $p_E=[-\mu~0~0]^\mr{T}$ and the position of the Moon is $p_M=[1-\mu~0~0]^\mr{T}$. Then, the rotating, non-dimensional equations of motion for the spacecraft in this frame are given by
\begin{equation}\label{eq:CR3BP}
\begin{split}
\ddot{x}&=2\dot{y}+x-\frac{(1-\mu)(x+\mu)}{r_\mr{e}^3}-\frac{\mu}{r_m^3}(x-1+\mu)+\frac{u_x}{m},\\
\ddot{y}&=-2\dot{x}+y-\frac{1-\mu}{r_\mr{e}^3}y-\frac{\mu}{r_m^3}y+\frac{u_y}{m},\\
\ddot{z}&=-\frac{1-\mu}{r_\mr{e}^3}z-\frac{\mu}{r_m^3}z+\frac{u_z}{m},
\end{split}
\end{equation}
where $(x,y,z)$ is the spacecraft's normalized position, $(\dot{x},\dot{y},\dot{z})$ its normalized velocity,   $(u_x,u_y,u_z)$ its normalized thrust, and $m$ its mass. In addition,
\begin{equation*}
r_\mr{e}=\sqrt{(x+\mu)^2+y^2+z^2},\qquad r_m=\sqrt{(x-1+\mu)^2+y^2+z^2},
\end{equation*}
denote the spacecraft's normalized distances from the Earth and the Moon, respectively.

We can also write the spacecraft equations of motion in compact form. Denoting $\mathbf{x}=[x ~ y ~ z ~ \dot{x} ~ \dot{y} ~ \dot{z}]^\mr{T}\in\mathbb{R}^6$ and $\mathbf{u}=[u_x ~ u_y ~ u_z]^\mr{T}\in\mathbb{R}^3$, such compact form is given by
\begin{equation}\label{eq:CR3BPc}
\dot{\mathbf{x}}=f(\mathbf{x})+B\mathbf{u},
\end{equation}
where $B = [0_{3} ~ \frac{1}{m}I_3 ]^\mr{T}$ and
\begin{equation}\label{eq:CR3BPc2}
\begin{split}
f(\mathbf{x}):=\begin{bmatrix}f_x(\mathbf{x}) \\ f_y(\mathbf{x}) \\f_z(\mathbf{x}) \\f_{\dot{x}}(\mathbf{x}) \\f_{\dot{y}}(\mathbf{x}) \\f_{\dot{z}}(\mathbf{x}) \end{bmatrix}=\begin{bmatrix}\dot{x} \\ \dot{y} \\ \dot{z} \\ 2\dot{y}+x-\frac{(1-\mu)(x+\mu)}{r_\mr{e}^3}-\frac{\mu}{r_m^3}(x-1+\mu) \\ -2\dot{x}+y-\frac{1-\mu}{r_\mr{e}^3}y-\frac{\mu}{r_m^3}y \\ -\frac{1-\mu}{r_\mr{e}^3}z-\frac{\mu}{r_m^3}z\end{bmatrix}.
\end{split}
\end{equation}

\subsection{Pursuit-Evasion in Cislunar Space}
We consider two spacecraft in cislunar space, the evader (e) and the pursuer (p), engaged in pursuit-evasion. Each spacecraft obeys the CR3BP dynamics of Section \ref{sec:CR3BP}, so that their states evolve according to
\begin{align*}
\dot{\mathbf{x}}_{\mathrm e}(t)&=f(\mathbf{x}_{\mathrm e}(t))+B_{\mathrm e}\mathbf{u}_{\mathrm e}(t),\qquad \mathbf{x}_{\mathrm e}(t_0)=\mathbf{x}_{\mathrm e0},\\
\dot{\mathbf{x}}_{\mathrm p}(t)&=f(\mathbf{x}_{\mathrm p}(t))+B_{\mathrm p}\mathbf{u}_{\mathrm p}(t),\qquad \mathbf{x}_{\mathrm p}(t_0)=\mathbf{x}_{\mathrm p0},
\end{align*}
where, for $\mr{i}\in\{\mathrm e,\mathrm p\}$, $\mathbf{x}_\mr{i}(t)$ denotes the state of spacecraft $\mr{i}$, $\mathbf{u}_\mr{i}(t)$ its thrust input, $B_\mr{i}=[0_3~ \frac{1}{m_\mr{i}}I_3]^\mr{T}$ its input matrix, and $m_\mr{i}$ its mass.

The evader aims to increase its distance from the pursuer, whereas the pursuer aims to minimize this distance. At the same time, both spacecraft seek to remain near a given reference orbit $\mf{x}_\mr{d}:\mathbb{T}^d\rightarrow\mathbb{R}^6$ in the vicinity of a Lagrange point, where $d=1$ if $\mf{x}_\mr{d}$ is a periodic orbit (PO) and $d=2$ if it is a quasi-periodic orbit (QPO)\footnote{We consider only two-dimensional quasi-periodic orbits. However, our theory can extend to higher-dimensional ones.}. Such trajectories are of practical interest because they provide useful operational geometries and known stability; we detail their construction and associated phasing dynamics in Section \ref{sec:phasing}.

\begin{remark}
Though it is more standard to define reference orbits as functions of time, here we define them over a \textit{torus} that can be one- or two-dimensional ($d=1$ or $d=2$). This convention allows us to represent both periodic and quasi-periodic reference orbits in a unified way, as we will see in subsequent sections.
\end{remark}

Since the spacecraft may have different phases along the reference orbit, let $\bm{\theta}_\mr{e}(t)\in\mathbb{T}^d$ and $\bm{\theta}_\mr{p}(t)\in\mathbb{T}^d$ denote the phases of the evader and pursuer, respectively. Then, their corresponding reference signals are
\begin{align*}
\mf{x}_\mr{d,e}(t)&=\mf{x}_\mr{d}(\bm{\theta}_\mr{e}(t)),\\
\mf{x}_\mr{d,p}(t)&=\mf{x}_\mr{d}(\bm{\theta}_\mr{p}(t)).
\end{align*}

In the remainder of the paper, we formulate and solve a differential game that captures the competing objectives of the two spacecraft, while allowing them to exploit the structure of the CR3BP to execute more aggressive maneuvers.

\section{Reference Construction and Phasing Adjustment}\label{sec:phasing}

The CR3BP exhibits chaotic dynamics, so pursuit-evasion maneuvers must be carefully designed to keep each spacecraft near the relative stability of its nominal orbit. However, restricting each spacecraft to a tight neighborhood of that orbit can significantly limit its maneuvering capability. To alleviate this tradeoff, we endow each spacecraft with additional decision variables beyond thrust, namely, the rates at which its reference phases evolve along the orbit. Such variables allow more aggressive pursuit or evasion without forcing the spacecraft to depart the geometric vicinity of the orbit. We develop them first for the PO case, and then for the QPO case, which affords considerably greater maneuvering flexibility.

\subsection{Periodic Orbits}

We first consider the case where the reference orbit is periodic, i.e., $d=1$ and $\mf{x}_\mr{d}:\mathbb{T}\rightarrow\mathbb{R}^6$ is a one-dimensional curve in state space. Such curves include, for example, Lyapunov, halo, and near-rectilinear halo orbits (NRHO), which can be computed via standard differential correction methods and are widely tabulated for cislunar missions \cite{ssd_periodic_orbits_2026}.

A natural restriction imposed by requiring the spacecraft to follow a specific periodic reference signal, i.e., $\mf{x}_\mr{d}(\bm{\theta}_\mr{e}(t))$ and $\mf{x}_\mr{d}(\bm{\theta}_\mr{p}(t))$, is that $\bm{\theta}_\mr{e}(t)$ and $\bm{\theta}_\mr{p}(t)$ are typically chosen to evolve according to physical time, i.e., $\dot{\bm{\theta}}_\mr{e}(t)=\dot{\bm{\theta}}_\mr{p}(t)=\frac{1}{T}$, where $\bm{\theta}_\mr{e},\bm{\theta}_\mr{p} \in [0,1]$, and $T$ is the orbital period. This choice guarantees that if $\mf{x}_\mr{d}$ is a solution of the CR3BP with zero thrust, then so are $\mf{x}_\mr{d,e}$ and $\mf{x}_\mr{d,p}$.
A fixed reference phase assigns the spacecraft a prescribed location on the reference at every time. The spacecraft must then use thrust to follow that moving point, even when another point on the same reference would be more favorable during the encounter. We instead allow each agent to control how fast its reference point moves. The physical spacecraft still follows the CR3BP dynamics, but its tracking target can move \textit{faster or slower} along the reference. Such flexibility enables along-track maneuvers without necessarily requiring the spacecraft to leave the reference geometry.

We therefore introduce the controlled reference phase dynamics of the spacecraft's reference orbit phasing:\footnote{We assume the phases are modulo 1. In other words, they reset to zero once they reach unity,  though without affecting the continuity of the corresponding reference signals.}
\begin{equation}\label{eq:genorbitphase}
\begin{split}
\dot{\bm{\theta}}_\mr{i}(t)&=\frac{1}{T}\bm{\omega}_\mr{i}(t), \quad \bm{\theta}_\mr{i}(0)=\bm{\theta}_\mr{0i},
\end{split}
\end{equation}
where $\mr{i}\in\{\mr{e},\mr{p}\},$ and $\bm{\omega}_\mr{i}(t)$ are phase control variables that scale the evolution of the reference orbit phasing. If $\bm{\omega}_\mr{i}(t)>1$, then the spacecraft complete their orbits faster than the nominal traversal time, i.e., they speed up along track. Conversely, if $\bm{\omega}_\mr{i}(t)<1$, then they slow down along the orbit. Hence, by adjusting $\bm{\omega}_\mr{i}(t)$, the spacecraft can control their reference position along the orbit.

\subsection{Quasi-Periodic Orbits}

\begin{figure}[t]
    \centering
    \begin{overpic}{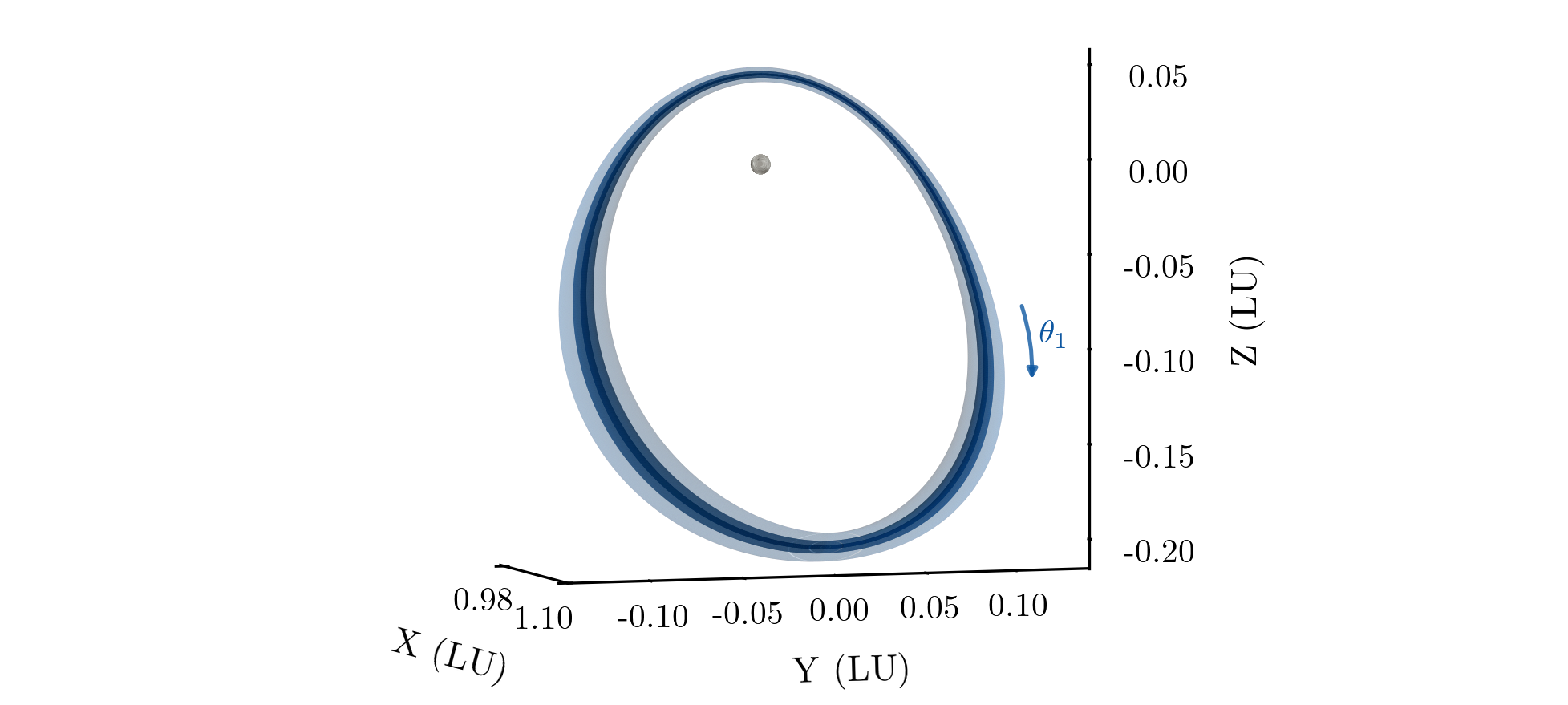}
        
        \put(0, 25){%
            \setlength{\fboxsep}{0pt}%
            \fbox{\includegraphics[width=0.3\textwidth]{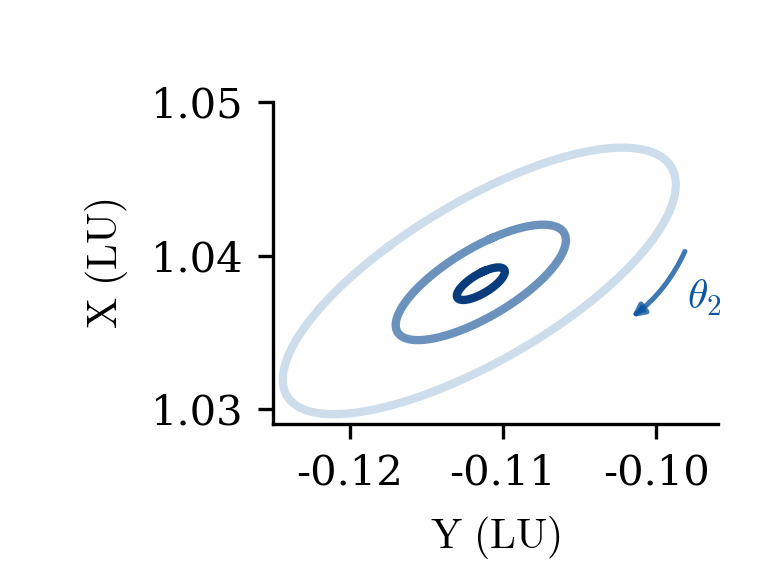}}
        }

        \put(35, 20){\color{black}\framebox(5, 5){}}
        \color{black}
        \thicklines 
        \Line(0, 25)(35, 20) 
        \Line(30.015, 47.49)(40, 25)

    \end{overpic}
    \caption{Representative family of $L_2$ southern quasi-halo orbits. The main 3D plot illustrates the invariant tori, where $\theta_1$ parametrizes the longitudinal direction. The zoomed-in detail window displays a 2D projection of a Poincaré section of the QPO family, with $\theta_2$ representing the transverse direction. The blue color gradient denotes successive members of the family generated via continuation.}
    \label{fig:QPO_family}
\end{figure}

Next, we consider the case where the reference orbit is quasi-periodic, i.e., $d=2$ and $\mf{x}_\mr{d}:\mathbb{T}^2\rightarrow\mathbb{R}^6$ parameterizes an invariant torus. Phase-rate control on a periodic orbit, as in the previous subsection, provides tracking flexibility along a one-dimensional curve. Quasi-periodic orbits (QPOs) extend this flexibility by allowing the spacecraft to maneuver over a two-dimensional invariant torus. Specifically, the torus is parameterized by two angles $\theta_1$ and $\theta_2$, that describe the quasi-periodic motion. The angle $\theta_1$ generalizes the phase variable along the longitudinal direction from the periodic case, while $\theta_2$ parametrizes the transverse direction on the torus, as illustrated in Fig.~\ref{fig:QPO_family}. By adjusting both phases, each spacecraft can steer its reference point to any point on the torus, rather than being restricted to a one-dimensional curve.

We construct an approximate Fourier representation of the two-dimensional torus, $\hat{K}\approx\mf{x}_\mr{d}$, and its stable/unstable bundle (i.e., the linear stable/unstable directions) in two steps. First, following the flow map parameterization method of Haro and Mondelo \cite{haro2021flow}, we compute a one-dimensional generator (curve) $K$ that satisfies a flow invariance condition. Second, we extend this by flowing the 1D generator to fit a full 2D Fourier series. The construction proceeds as follows.

\subsubsection{One-Dimensional Generator of the Torus}

Let $K:\mathbb{T}\rightarrow\mathbb{R}^6$ denote a one-dimensional generator corresponding to a torus Poincaré section, and let $W:\mathbb{T}\rightarrow\mathbb{R}^6$ be the corresponding stable or unstable bundle. The generator $K$ and the bundle $W$ satisfy the invariance equations
\begin{align}\label{eq:invariance}
    \varphi_T(K(\theta))&=K(\theta+\rho),\\ 
    D\varphi_T(K(\theta))W(\theta)&=\lambda W(\theta+\rho),~\forall\theta\in\mathbb{T},\label{eq:invariance2}
\end{align}
where $\varphi_t$ denotes the Hamiltonian flow of the CR3BP dynamics, $D\varphi_t$ is the Jacobian of the flow, $T\in\mathbb{R}$ is the time of flight (or ``period''), $\lambda\in\mathbb{R}$ is the bundle $W$ multiplier over one period, and $\rho\in[0,1]$ is the rotation number (i.e., the trajectory rotates by $2\pi\rho$ rad per period). In other words, if any point on the generator curve is propagated forward by one period $T$ under the CR3BP flow, it returns to the same invariant curve on the Poincaré section but shifted by the rotation number $\rho$. 

To compute $K$, we enforce \eqref{eq:invariance} and \eqref{eq:invariance2} numerically using a multi-shooting scheme. The method begins with an initial guess for $m$ segments $\{K_i\}_{i=1}^m$ along the generator and the corresponding multiple bundles $\{W_i\}_{i=1}^m$. We then correct this initial guess via Newton iterations. This correction step relies on transforming the system of linearized invariance equations into a block upper-triangular form via an adapted symplectic frame. However, in practice, disparate column magnitudes within this frame can lead to an ill-conditioned Newton system. To resolve this, we introduce a uniform normalization scheme that builds upon \cite[Remark 3.3.1]{haro2021flow}. Specifically, we apply a constant linear transformation to the tangent block (representing the 2D space spanned by the longitudinal and transverse directions), derived via the Cholesky factorization of the average Gram matrix, alongside a scalar normalization to the bundle direction. This enforces an average orthonormal basis within the tangent block and a unit average norm for the bundle direction, while preserving the invariant subspaces (see Appendix~\ref{app:Frame} for the full derivation). Once an initial invariant torus is computed, we generate subsequent members of the same family by performing numerical continuation with respect to the torus period.

\subsubsection{Two-Dimensional Quasi-Periodic Parameterization}

Once the one-dimensional generator $K$ is known, the full two-dimensional torus, denoted $\hat{K}:\mathbb{T}^2\rightarrow\mathbb{R}^6$, can be recovered by flowing this generator in time and fitting a Fourier series. The construction relies on the following property of the flow: propagating a point $K(\theta_2)$ on the generator forward by time $t\in[0,T]$ yields a new point on the torus at angle $\theta_1=t/T$ and at angle $\theta_2$ shifted by $(t/T)\rho$, i.e.,
\begin{align*}
\varphi_t(K(\theta_2))
=
\hat{K}\!\left(\tfrac{t}{T},\,\theta_2+\tfrac{t}{T}\rho\right).
\end{align*}
Thus, varying $t$ from $0$ to $T$ sweeps the $\theta_1$ angle across $[0,1]$, while the $\theta_2$ argument simultaneously drifts. We build a uniform two-dimensional grid by discretizing $\theta_1$ with $m$ evenly spaced nodes $\theta_{1,i}=i/m$ for $i=0,\dots,m-1$, and $\theta_2$ with $N$ uniform nodes $\theta_{2,j}$ for $j=0,\dots,N-1$. Pre-shifting the generator argument by $(i/m)\rho$ to compensate for the drift, we obtain the grid
\begin{align*}
    K_{i,j}
    =
    \varphi_{\frac{i}{m}T}
    \left(
    K\left(\theta_{2,j}-\frac{i}{m}\rho\right)
    \right),
    \quad
    i=0,\dots,m-1,\quad j=0,\dots,N-1,
\end{align*}
which gives torus samples at the angular pairs $(\theta_{1,i},\theta_{2,j})$. From these samples, we obtain a truncated Fourier series representation of the torus:
\begin{align}
    \hat{K}(\theta_1,\theta_2)
    =
    \sum_{k=-\lfloor m/2\rfloor}^{\lceil m/2\rceil-1}
    \sum_{l=-\lfloor N/2\rfloor}^{\lceil N/2\rceil-1}
    \widetilde{K}_{k,l}
    e^{\mf{i}2\pi(k\theta_1+l\theta_2)},
\end{align}
where $\widetilde{K}_{k,l}\in\mathbb{C}^6$ are the Fourier coefficients obtained from a two-dimensional discrete Fourier transform.

Finally, given the two-parameter torus parameterization, we identify the reference orbit with its Fourier representation, $\mf{x}_\mr{d}(\bm{\theta}):=\hat{K}(\bm{\theta})$, and define for each spacecraft the phase tuple $\bm{\theta}_\mr{i}=(\theta_{\mr{i},1},\theta_{\mr{i},2})\in\mathbb{T}^2$, $\mr{i}\in\{\mr{e},\mr{p}\}$, so that the reference signal becomes $\mf{x}_{\mr{d}i}(t)=\mf{x}_\mr{d}(\bm{\theta}_\mr{i}(t))$. The agents then control their reference phasing according to
\begin{align}\label{eq:qpoorbitphase}
\dot{\theta}_{\mr{i},1}(t)
&=
\frac{1}{T}\omega_{\mr{i},1}(t),
\quad
\theta_{\mr{i},1}(t_0)=\theta_{01,\mr{i}},\\
\dot{\theta}_{\mr{i},2}(t)
&=
\frac{\rho}{T}\omega_{\mr{i},2}(t),
\quad
\theta_{\mr{i},2}(t_0)=\theta_{02,\mr{i}},
\end{align}
where $\bm{\omega}_\mr{i}(t)=(\omega_{\mr{i},1}(t),\omega_{\mr{i},2}(t))\in\mathbb{R}^2$ contains the phase-rate controls along the longitudinal and transverse directions of the torus, respectively. This is similar to the PO case \eqref{eq:genorbitphase}, but with two controllable phases ($d=2$) rather than one, which broadens the spacecraft's maneuvering capabilities.

\begin{remark}
Note that, once the torus is represented in Fourier form, derivatives with respect to the angular variables are computed directly in Fourier space. In particular, the first- and second-order derivatives are
\begin{align}
    \widetilde{(\partial_{\theta_j}\hat{K})}_{\mathbf{k}}
    &=
    \mf{i}2\pi k_j\,\widetilde{K}_{\mathbf{k}},\\
    \widetilde{(\partial_{\theta_i}\partial_{\theta_j}\hat{K})}_{\mathbf{k}}
    &=
    (\mf{i}2\pi k_i)(\mf{i}2\pi k_j)\,\widetilde{K}_{\mathbf{k}},
\end{align}
where $\mathbf{k}=(k_1,k_2)\in\mathbb{Z}^2$ and $i,j\in\{1,2\}$.
\end{remark}

\section{Zero-Sum Dynamic Game for Adversarial Pursuits in Cislunar Space}\label{sec:CGame}

Equipped with the phasing dynamics of Section \ref{sec:phasing}, we now formulate the pursuit-evasion differential game between the two spacecraft. Subsequently, we propose an algorithm that approximates a saddle-point solution to this game by i) adaptively discretizing the game in time and ii) employing iterative linear-quadratic approximations.

Recall that the evader and the pursuer aim to increase and decrease, respectively, the distance between them, all while tracking the phase-controllable reference orbit. To do so, each spacecraft uses its thrust $\mf{u}_\mr{i}$ to control its position, and its phase-rate control $\bm{\omega}_\mr{i}$ to control its reference signal. We collect both into the full decision vector 
$\mf{w}_\mr{i} := [\mf{u}_\mr{i}^\mr{T} ~ \bm{\omega}_\mr{i}^\mr{T}]^\mr{T}$
and define the concatenated state
\begin{equation*}
\mf{x} := [\mf{x}_\mr{e}^\mr{T} ~ \mf{x}_\mr{p}^\mr{T} ~ \bm{\theta}_\mr{e}^\mr{T} ~ \bm{\theta}_\mr{p}^\mr{T}]^\mr{T} \in \mathbb{R}^{12}\times\mathbb{T}^{2d}.
\end{equation*}
 Denoting also the nominal phase evolution as $\bm{\Omega}_{0} \in \mathbb{R}^d$ (e.g., $\bm{\Omega}_{0,\text{PO}} = \frac{1}{T}$, $\bm{\Omega}_{0,\text{QPO}} = [\frac{1}{T}, \frac{\rho}{T}]^\mr{T}$), 
the resulting zero-sum differential game is
\begin{align}\label{eq:nlgame}
\min_{\mathbf{w}_\mr{e}\in\mathcal{W}_\mr{e}} \max_{\mathbf{w}_\mr{p}\in\mathcal{W}_\mr{p}} \int_{t_0}^{t_f} L(\mathbf{x}(t),\mathbf{w}_\mr{e}(t),\mathbf{w}_\mr{p}(t),t)\mathrm{d}t + \varphi(\mathbf{x}(t_f), t_f),
\end{align}
subject to the dynamics 
\begin{align}\label{eq:nonlineardyn}
\dot{\mf{x}}(t)=F(\mf{x},\mf{w}_\mr{e},\mf{w}_\mr{p}):=\begin{bmatrix} 
f({\mathbf{x}}_\mathrm{e}(t))+B_\mr{e}\mathbf{u}_\mathrm{e}(t)  \\
f({\mathbf{x}}_\mathrm{p}(t))+B_\mr{p}\mathbf{u}_\mathrm{p}(t) \\ \bm{\omega}_\mr{e}(t)\circ\bm{\Omega}_{0} \\ 
\bm{\omega}_\mr{p}(t)\circ\bm{\Omega}_{0}\end{bmatrix}, \quad\mf{x}(t_0)=\mf{x}_0:=\begin{bmatrix}{\mathbf{x}}_\mathrm{e0}  \\  {\mathbf{x}}_\mathrm{p0} \\ \bm{\theta}_\mr{0e}  \\ \bm{\theta}_\mr{0p} \end{bmatrix},
\end{align}
and the input constraints $\mathbf{w}_\mr{e}\in\mathcal{W}_\mr{e}$, $\mathbf{w}_\mr{p}\in\mathcal{W}_\mr{p}$, given by
\begin{align*}
\mf{u}^-_\mr{e}
\leq\mf{u}_\mr{e}(t)\leq\mf{u}^+_\mr{e},\qquad
\bm{\omega}^-_\mr{e}\leq\bm{\omega}_\mr{e}(t)\leq\bm{\omega}^+_\mr{e},\\
\mf{u}^-_\mr{p} \leq \mf{u}_\mr{p}(t) \leq \mf{u}^+_\mr{p}, \qquad 
\bm{\omega}^-_\mr{p} \leq \bm{\omega}_\mr{p}(t) \leq \bm{\omega}^+_\mr{p}.
\end{align*}
The running and terminal costs of the game decompose as 
\begin{align*}
L &:= \norm{\Delta{\mathbf{x}}_\mathrm{e}(t)}_{Q_\mr{e}(t)}^2 + \norm{\mathbf{u}_\mathrm{e}(t)}_{R_\mathrm{e}(t)}^2 + \norm{\bm{\omega}_\mr{e}(t) - \mf{1}_\mr{d}}_{A_\mr{e}(t)}^2 \\
&\quad - \norm{\Delta{\mathbf{x}}_\mathrm{p}(t)}_{Q_\mathrm{p}(t)}^2 - \norm{\mathbf{u}_\mathrm{p}(t)}_{R_\mathrm{p}(t)}^2 - \norm{\bm{\omega}_\mr{p}(t) - \mf{1}_\mr{d}}_{A_\mr{p}(t)}^2 \\
&\quad + S\left(\norm{\mf{r}_{\mr{e}}(t)-\mf{r}_{\mr{p}}(t)}\right), \\
\phi &:= \norm{\Delta{\mathbf{x}}_\mathrm{e}(t_f)}_{F_\mathrm{e}}^2 - \norm{\Delta{\mathbf{x}}_\mathrm{p}(t_f)}_{F_\mr{p}}^2 + S\left(\norm{\mf{r}_{\mr{e}}(t_f)-\mf{r}_{\mr{p}}(t_f)}\right),
\end{align*}
where $\Delta\mathbf{x}_\mr{i}(t) = \mathbf{x}_\mr{i}(t) - \mathbf{x}_\mr{d}({\bm{\theta}}_\mr{i}(t))$ denotes the tracking error of player $\mr{i}$ relative to the phase-parameterized reference orbit, and $\bm{\omega}_\mr{i}(t) \in \mathbb{R}^{d}$ represents the vector of reference phase rate controls for agent $\mr{i} \in \{\mr{e}, \mr{p}\}$.  The weighting matrices $Q_\mr{i},~ F_\mr{i},~ R_\mr{i},~A_\mr{i} \succ 0$ penalize, respectively, running and terminal tracking deviations, control effort, and phase-rate deviations from nominal values. Finally, $S: \mathbb{R}_+ \rightarrow \mathbb{R}_+$  is a function that penalizes the evader for being close to the pursuer, and vice versa for the pursuer. A relevant choice for this function is:
\begin{equation}\label{eq:evadecost}
S(D)=\begin{cases}\frac{1}{p}w(D_0-D)^p,~ &D\le D_0 \\ 0   ~&D\ge D_0 \end{cases},
\end{equation}
where $w>0$, $p>2$, $D_0>0$, with $D_0$ being the escape distance and $D=\norm{\mf{r}_{\mr{e}}(t)-\mf{r}_{\mr{p}}(t)}_2$. Note that this function $S$ is twice continuously differentiable; a property that we will find useful when employing numerical tools to solve the differential game. 

\section{Adaptive Discretization for Multi-Agent Trajectory Optimization}

The game in the previous section has a non-quadratic cost, nonlinear dynamics, and input constraints, making it difficult to solve analytically. We therefore resort to algorithmic methods to solve it approximately.

One relevant algorithmic method is continuous-time DDP, which converts the game into a sequence of linear-quadratic problems solved iteratively \cite{sun2018min}. However, continuous-time DDP can be computationally expensive as it requires repeated integration of differential Riccati equations, and becomes even more challenging in the presence of constraints. 

An alternative is to discretize the game and apply a discrete-time version of DDP to solve it. This approach works well for halo and Lyapunov orbits under sufficiently fine discretization. However, for near-rectilinear halo orbits that pass close to the Moon, discretization errors can grow significantly, requiring prohibitively fine time steps. In addition, the rapid variation of the gravitational field during close approaches leads to poorly conditioned and rapidly varying dynamics, making the game more difficult to solve reliably. To address these challenges, we combine a discrete-time DDP formulation with adaptive time discretization using multi-agent Sundman transformation and Kustaanheimo-Stiefel (KS) regularization.

\subsection{Coordinate Regularization}\label{sec:KS}

We regularize the CR3BP dynamics \eqref{eq:CR3BP} for each spacecraft using the KS transformation centered at the Moon \cite{howell1984almost}. The KS transformation maps the 3D Cartesian coordinates into a fictitious 4D space, weakening the gravitational singularity at the Moon from a $1/r^2$ dependence to a $1/r$ term. The remaining $1/r$ singularity is then fully removed by a Sundman time transformation, $dt = r\,d\tau$, which rescales physical time according to the spacecraft's distance $r$ from the Moon. Applied independently to the evader and the pursuer, this yields a fictitious time per spacecraft: $\tau_\mr{e}$ governed by $dt = r_\mr{e}\,d\tau_\mr{e}$, and $\tau_\mr{p}$ governed by $dt = r_\mr{p}\,d\tau_\mr{p}$, where $r_\mr{e}$ and $r_\mr{p}$ are the respective lunar distances. See Appendix~\ref{app:KS} and \cite{howell1984almost, stiefel1971linear} for details. 

In the regularized space, each spacecraft's state becomes ten-dimensional,
\begin{equation*}
\mathbf{y}_\mr{i} = [\mathbf{z}_\mr{i}^\mr{T},\; \mathbf{z}_\mr{i}^{'\mr{T}},\; H_\mr{i},\; t]^\mr{T} \in \mathbb{R}^{10},\qquad i \in \{\mr{e},\mr{p}\},
\end{equation*}
where $\mathbf{z}_\mr{i} \in \mathbb{R}^4$ is the KS position, $\mathbf{z}_\mr{i}' \in \mathbb{R}^4$ is the velocity with respect to $\tau_\mr{i}$, $H_\mr{i} \in \mathbb{R}$ is the CR3BP Hamiltonian, and $t$ is the physical time, carried along as part of the state. The per-spacecraft regularized dynamics take the form
\begin{equation}\label{eq:KS_single}
\frac{d\mathbf{y}_\mr{i}}{d\tau_\mr{i}} = f_{KS}(\mathbf{y}_\mr{i}) + B_{KS,i}(\mathbf{y}_\mr{i})\,\mathbf{u}_\mr{i},
\end{equation}
with $f_{KS}$ and $B_{KS,i}$ described in Appendix~\ref{app:KS}.
The transformation back to Cartesian coordinates is 
\begin{equation}\label{eq:conv_KS_cart}
\mathbf{x}_\mr{i} := \kappa(\mathbf{y}_\mr{i}) = \begin{bmatrix} \mathbf{r}_\mr{i} \\ \mathbf{v}_\mr{i} \end{bmatrix} = \begin{bmatrix} L(\mathbf{z}_\mr{i})\mathbf{z}_\mr{i} \\ \dfrac{2}{z_\mr{i}^2} L(\mathbf{z}_\mr{i})\mathbf{z}_\mr{i}' \end{bmatrix},
\end{equation}
where $L(\mathbf{z}_\mr{i})$ is the KS matrix (also defined in Appendix~\ref{app:KS}). The two player conversion map is then $\mf{x} := \kappa(\mf{y}):=[\kappa(\mf{y}_\mr{e})^\mr{T}~\kappa(\mf{y}_\mr{p})^\mr{T}~\bm{\theta}_\mr{e}^\mr{T}~\bm{\theta}_\mr{p}^\mr{T}]^\mr{T}$. This mapping allows the running and terminal costs of the game \eqref{eq:nlgame} to be evaluated using the regularized state $\mathbf{y}$.

One issue remains: the per-spacecraft fictitious times $\tau_\mr{e}$ and $\tau_\mr{p}$ are decoupled. A uniform step in $\tau_\mr{e}$ does not correspond to a uniform step in $\tau_\mr{p}$, since the two transformations are tied to different lunar distances. We need to evaluate the zero-sum cost, which depends on the joint state at a common physical time. Therefore it requires synchronizing the two clocks. We address this point in the next subsection by introducing a single global fictitious time $s$.

\subsection{Multi-Agent Time Regularization and Synchronization}
\label{sec:discrete_time}

The single-spacecraft Sundman transformation of Section \ref{sec:KS} regularizes the gravitational singularity at the Moon by introducing a fictitious time whose rate is tied to the spacecraft's distance from the Moon. Applied separately to the evader and the pursuer, however, it yields \emph{two} fictitious times, $\tau_\mr{e}$ and $\tau_\mr{p}$ and physical times $t_\mr{e}$ and $t_\mr{p}$, governed by $dt_\mr{e} = r_\mr{e}\,d\tau_\mr{e}$ and $dt_\mr{p} = r_\mr{p}\,d\tau_\mr{p}$, where $r_\mr{e}$ and $r_\mr{p}$ denote the evader's and pursuer's respective distances from the Moon. Because in general $r_\mr{e}\neq r_\mr{p}$, uniform stepping in $\tau_\mr{e}$ doesn't correspond to a uniform stepping in $\tau_\mr{p}$. Two agents evaluated at the same discretization step exist at different physical times, as if each spacecraft were interacting with its own independent copy of the Moon. An agent can exploit this by driving its lunar distance toward zero, effectively halting its own progression in physical time and gaining an unphysical advantage in evasion or interception.

Eliminating this exploit requires a single fictitious time $s$ that advances physical time identically for both spacecraft. That is, $s$ must satisfy
\begin{equation}
    f_s(r_e,r_p) ds = r_\mr{e}\,d\tau_\mr{e} = r_\mr{p}\,d\tau_\mr{p} = dt,
\end{equation}
for some  shared time-scaling function $f_s(r_e,r_p)$. Writing $d\tau_\mr{e} = \alpha_\mr{e}\,ds$ and $d\tau_\mr{p} = \alpha_\mr{p}\,ds$ for some scalar factors $\alpha_\mr{e}, \alpha_\mr{p}$, this requirement gives $r_\mr{e}\alpha_\mr{e} = r_\mr{p}\alpha_\mr{p}$, which (up to a normalization choice) yields
\begin{equation}\label{eq:sync_func}
    d\tau_\mr{e} = \frac{r_\mr{p}}{r_\mr{e}+r_\mr{p}}\,ds, \qquad
    d\tau_\mr{p} = \frac{r_\mr{e}}{r_\mr{e}+r_\mr{p}}\,ds, \qquad
    dt = \frac{r_\mr{e}\,r_\mr{p}}{r_\mr{e}+r_\mr{p}}\,ds.
\end{equation}
Equivalently, $dt$ scales inversely with the sum of inverse lunar distances, i.e., $dt = (1/r_\mr{e} + 1/r_\mr{p})^{-1}\,ds$ and hence $f_s(r_e,r_p)=(1/r_\mr{e} + 1/r_\mr{p})^{-1}$. This construction parallels the global N-body regularization of Heggie~\cite{heggie1974global}, who introduced a shared fictitious time to handle simultaneous near-collisions among multiple bodies. Here we apply the same principle for a different purpose: not to handle multiple simultaneous singularities, but to maintain temporal synchronization between independent agents under a single regularized clock.

The shared time $s$ acts as a variable-rate clock for the coupled system. When both spacecraft are far from the Moon, $dt/ds$ is large and the effective step in physical time remains roughly uniform. As either spacecraft approaches the Moon, $dt/ds$ shrinks, increasing temporal resolution for the joint system; the distant spacecraft is correspondingly slowed in its own fictitious time $\tau_\mr{i}$ to maintain synchronization. The result is a discretization that concentrates resolution where the joint trajectory approaches the singularity, without the inconsistencies of independent Sundman parameterizations.

Discretizing the coupled system uniformly in $s$ with step $\Delta s$, and using \eqref{eq:sync_func} to convert into each spacecraft's local Sundman increment $\Delta\tau_\mr{i}$, yields the joint dynamics
\begin{align}\label{eq:discrete_coupled}
\mathbf{y}(k+1) = \mathbf{y}(k) + \frac{\Delta s}{r_\mr{e}(k)+r_\mr{p}(k)}
\begin{bmatrix}
    r_\mr{p}(k)\,\bigl(f_{KS}(\mathbf{y}_\mr{e}(k)) + B_{KS,\mr{e}}(\mathbf{y}_\mr{e}(k))\mathbf{u}_\mr{e}(k)\bigr) \\
    r_\mr{e}(k)\,\bigl(f_{KS}(\mathbf{y}_\mr{p}(k)) + B_{KS,\mr{p}}(\mathbf{y}_\mr{p}(k))\mathbf{u}_\mr{p}(k)\bigr) \\
    r_\mr{e}(k)\,r_\mr{p}(k)\,\bm{\omega}_\mr{e}(k)\circ\bm{\Omega}_{0} \\ 
    r_\mr{e}(k)\,r_\mr{p}(k)\,\bm{\omega}_\mr{p}(k)\circ\bm{\Omega}_{0}
\end{bmatrix},
\end{align}
where $k \in \{0, 1, \ldots, N-1\}$ is the discrete index in $s$ and $\Delta\tau_\mr{e}(k) = \Delta s\,r_\mr{p}(k)/(r_\mr{e}(k)+r_\mr{p}(k))$, $\Delta\tau_\mr{p}(k) = \Delta s\,r_\mr{e}(k)/(r_\mr{e}(k)+r_\mr{p}(k))$ are the corresponding local Sundman increments. 

In practice, the shared time scaling must be estimated by both agents. Each spacecraft obtains its own lunar distance from onboard navigation and the other spacecraft’s distance from relative measurements or ground-based tracking. Errors in the scaling estimate affect the placement of the discretization nodes rather than the physical dynamics and can be reduced through repeated state estimation and replanning.

\section{Algorithmic Approximate Solution of Pursuit-Evasion Game}
In this section, we discretize and solve the pursuit-evasion game using the adaptive discretization of Section~\ref{sec:discrete_time}.

\subsection{Discretized Pursuit-Evasion Game}
Using the discretized dynamics \eqref{eq:discrete_coupled} of Section~\ref{sec:discrete_time}, we formulate the pursuit-evasion game of Section~\ref{sec:CGame} in discrete time as
\begin{equation}\label{eq:nlgame_discrete}
\min_{\{\mathbf{w}_{\mr{e}}(k)\}\in\mathcal{W}_\mr{e}^N}\max_{\{\mathbf{w}_{\mr{p}}(k)\}\in\mathcal{W}_\mr{p}^N} J(\{\mathbf{w}_{\mr{e}}(k)\},\{\mathbf{w}_{\mr{p}}(k)\})=\sum_{k=0}^{N-1}\mathcal{L}(\kappa(\mathbf{y}(k)),\mathbf{w}_{\mr{e}}(k),\mathbf{w}_{\mr{p}}(k),t(k))+\phi(\kappa(\mathbf{y}(N)),t(N)),
\end{equation}
where $\{\cdot\}$ denotes concatenation over time and
\begin{equation*}
\mathcal{L}(\kappa(\mathbf{y}(k)),\mathbf{w}_{\mr{e}}(k),\mathbf{w}_{\mr{p}}(k),t(k)):=L(\kappa(\mathbf{y}(k)),\mathbf{w}_{\mr{e}}(k),\mathbf{w}_{\mr{p}}(k),t(k))\frac{r_\mr{e}(k)r_\mr{p}(k)}{r_\mr{e}(k)+r_\mr{p}(k)}\Delta s
\end{equation*}
is the discretized cost integrand. Here, the factor $\frac{r_\mr{e}(k)r_\mr{p}(k)}{r_\mr{e}(k)+r_\mr{p}(k)}\Delta s$ is the physical-time increment $\Delta t(k)$ in \eqref{eq:discrete_coupled}, corresponding to one step of $\Delta s$ and converting $\int L\,dt$ into a Riemann sum in $s$, whereas $t(k)=t_0+\sum_{i=0}^{k-1}\Delta t(i)$. In addition, we used the mapping \eqref{eq:conv_KS_cart} to express $\mathbf{x}$ in terms of $\mathbf{y}$.
The game is subject to the dynamics \eqref{eq:discrete_coupled} and the input constraints
\begin{align}
\begin{split}\label{eq:input_con}
&\mf{u}^-_\mr{e}
\leq\mf{u}_\mr{e}(k)\leq\mf{u}^+_\mr{e},\qquad
\bm{\omega}^-_\mr{e}\leq\bm{\omega}_\mr{e}(k)\leq\bm{\omega}^+_\mr{e},\\
&\mf{u}^-_\mr{p} \leq \mf{u}_\mr{p}(k) \leq \mf{u}^+_\mr{p}, \qquad 
\bm{\omega}^-_\mr{p} \leq \bm{\omega}_\mr{p}(k) \leq \bm{\omega}^+_\mr{p}, \quad \forall k.
\end{split}
\end{align}
We seek a saddle-point solution $(\{\mathbf{w}_{\mr{e}}(k)\}^\star,\{\mathbf{w}_{\mr{p}}(k)\}^\star)$ to this game, i.e., a tuple of policies satisfying
\begin{equation*}
J(\{\mathbf{w}_{\mr{e}}(k)\}^\star,\{\mathbf{w}_{\mr{p}}(k)\})\le J(\{\mathbf{w}_{\mr{e}}(k)\}^\star,\{\mathbf{w}_{\mr{p}}(k)\}^\star)\le J(\{\mathbf{w}_{\mr{e}}(k)\},\{\mathbf{w}_{\mr{p}}(k)\}^\star),\quad\forall~\{\mathbf{w}_\mr{e}\},\{\mathbf{w}_\mr{p}\}.
\end{equation*}
Such a saddle-point solution can be characterized through the value function $V:\mathbb{R}^{20}\times\mathbb{T}^{2d}\times[0,N]\rightarrow\mathbb{R}$, which satisfies the discrete-time Hamilton--Jacobi--Isaacs (HJI) equation
\begin{equation}\label{eq:HJI}
\begin{split}
V(\mathbf{y}(k),k)&=\min_{\mathbf{w}_{\mr{e}}(k)\in\mathcal{W}_\mr{e}}\max_{\mathbf{w}_{\mr{p}}(k)\in\mathcal{W}_\mr{p}}\big\{\mathcal{L}(\kappa(\mathbf{y}(k)),\mathbf{w}_{\mr{e}}(k),\mathbf{w}_{\mr{p}}(k),t(k))+V(\mathbf{y}(k+1),k+1)\big\},\\ 
V(\mathbf{y}(N),N)&=\phi(\kappa(\mathbf{y}(N)),t(N)).
\end{split}
\end{equation}
However, solving the HJI equation analytically is intractable in general. For this reason,  we approximate a saddle-point solution of the game by applying a discrete-time DDP method. Thanks to the adaptive time discretization, the resulting discrete-time problem achieves an effective tradeoff between approximation accuracy and computational complexity by refining the discretization only where needed.

\subsection{Discrete-Time DDP}

\subsubsection{Local Linear-Quadratic Approximation}
DDP works by approximating the solution to \eqref{eq:HJI} through iterative linear-quadratic problems. At each iteration, it performs an expansion of \eqref{eq:HJI} about the current iterate $(\bar{\mathbf{y}},~\bar{\mathbf{w}}_\mr{e},~\bar{\mathbf{w}}_\mr{p})$, which is linear in the CR3BP dynamics and quadratic in the value function. We denote variables evaluated at the current iterate with a bar, e.g., $\bar{V}=V(\bar{\mathbf{y}}(k),k)$, and define the perturbations
\begin{equation}\label{eq:forward}
\delta\mathbf{y}=\mathbf{y}-\bar{\mathbf{y}}, \quad 
\delta\mathbf{w}_{\mr{e}}=\mathbf{w}_{\mr{e}}-\bar{\mathbf{w}}_{\mr{e}}, \quad 
\delta\mathbf{w}_{\mr{p}}=\mathbf{w}_{\mr{p}}-\bar{\mathbf{w}}_{\mr{p}}.
\end{equation}
Letting $\Phi$ denote the discrete dynamics map of \eqref{eq:discrete_coupled}, i.e., $\mathbf{y}(k+1) = \Phi(\mathbf{y}(k), \mathbf{w}_\mr{e}(k), \mathbf{w}_\mr{p}(k))$, a first-order approximation about the current iterate is
\begin{equation}\label{eq:lin_disc_dyn}
\delta \mf{y}(k+1)\approx \bar{\Phi}_{\mf{y}}\delta\mf{y}(k)+\bar{\Phi}_{\mf{w}_{\mr{e}}}\delta\mf{w}_{\mr{e}}(k)+\bar{\Phi}_{\mf{w}_{\mr{p}}}\delta\mf{w}_{\mr{p}}(k).
\end{equation}
In addition, letting $\Theta(\mathbf{y}(k),\mathbf{w}_{\mr{e}}(k),\mathbf{w}_{\mr{p}}(k),k):=\mathcal{L}(\kappa(\mathbf{y}(k)),\mathbf{w}_{\mr{e}}(k),\mathbf{w}_{\mr{p}}(k),t(k))+V(\mathbf{y}(k+1),k+1)$, a second-order expansion of the right-hand side of \eqref{eq:HJI} about the current iterate is
\begin{multline*}
\min_{\mathbf{w}_{\mr{e}}(k)\in\mathcal{W}_\mr{e}}\max_{\mathbf{w}_{\mr{p}}(k)\in\mathcal{W}_\mr{p}}\big\{\mathcal{L}(\kappa(\mathbf{y}(k)),\mathbf{w}_{\mr{e}}(k),\mathbf{w}_{\mr{p}}(k),t(k))+V(\mathbf{y}(k+1),k+1)\big\} \\
\approx \min_{\delta\mathbf{w}_{\mr{e}}(k)\in\delta\mathcal{W}_\mr{e}}\max_{\delta\mathbf{w}_{\mr{p}}(k)\in\delta\mathcal{W}_\mr{p}}\left\{\bar{\Theta}+\delta\mf{y}(k)^\mr{T}\bar{\Theta}_\mf{y}+\delta\mf{w}_{\mr{e}}(k)^\mr{T}\bar{\Theta}_{\mf{w}_\mr{e}}+\delta\mf{w}_{\mr{p}}(k)^\mr{T}\bar{\Theta}_{\mf{w}_\mr{p}}\right.\\
\left.+\frac{1}{2}\begin{bmatrix}\delta\mf{y}(k) \\ \delta\mf{w}_{\mr{e}}(k) \\ \delta\mf{w}_{\mr{p}}(k) \end{bmatrix}^\mr{T} \begin{bmatrix}\bar{\Theta}_{\mf{y}\mf{y}} & \bar{\Theta}_{\mf{y}\mf{w}_{\mr{e}}} & \bar{\Theta}_{\mf{y}\mf{w}_{\mr{p}}} \\ \bar{\Theta}_{\mf{w}_{\mr{e}}\mf{y}} & \bar{\Theta}_{\mf{w}_{\mr{e}}\mf{w}_{\mr{e}}} & \bar{\Theta}_{\mf{w}_{\mr{e}}\mf{w}_{\mr{p}}} \\ \bar{\Theta}_{\mf{w}_{\mr{p}}\mf{y}} & \bar{\Theta}_{\mf{w}_{\mr{p}}\mf{w}_{\mr{e}}} & \bar{\Theta}_{\mf{w}_{\mr{p}}\mf{w}_{\mr{p}}}\end{bmatrix}\begin{bmatrix}\delta\mf{y}(k) \\ \delta\mf{w}_{\mr{e}}(k) \\ \delta\mf{w}_{\mr{p}}(k) \end{bmatrix}\right\},
\end{multline*}
where
\begin{align*}
\bar{\Theta}_\mf{y} &= \bar{\Phi}_\mf{y}(k)^\mr{T}  \bar{V}_\mf{y}(k+1) + \bar{\mathcal{L}}_\mf{y}(k), \\
\bar{\Theta}_{\mf{w}_\mr{e}} &= \bar{\Phi}_{\mf{w}_\mr{e}}(k)^\mr{T} \bar{V}_\mf{y}(k+1) + \bar{\mathcal{L}}_{\mf{w}_\mr{e}}(k), \\
\bar{\Theta}_{\mf{w}_\mr{p}} &= \bar{\Phi}_{\mf{w}_\mr{p}}(k)^\mr{T} \bar{V}_\mf{y}(k+1) + \bar{\mathcal{L}}_{\mf{w}_\mr{p}}(k), \\
\bar{\Theta}_{\mf{yy}} &= \bar{\Phi}_\mf{y}(k)^\mr{T} \bar{V}_{\mf{yy}}(k+1) \bar{\Phi}_\mf{y}(k) + \bar{\mathcal{L}}_{\mf{yy}}(k), \\
\bar{\Theta}_{\mf{w}_\mr{e}\mf{w}_\mr{e}} &= \bar{\Phi}_{\mf{w}_\mr{e}}(k)^\mr{T} \bar{V}_{\mf{yy}}(k+1) \bar{\Phi}_{\mf{w}_\mr{e}}(k) + \bar{\mathcal{L}}_{\mf{w}_\mr{e}\mf{w}_\mr{e}}(k), \\
\bar{\Theta}_{\mf{w}_\mr{e}\mf{w}_\mr{p}} &= \bar{\Phi}_{\mf{w}_\mr{e}}(k)^\mr{T} \bar{V}_{\mf{yy}}(k+1) \bar{\Phi}_{\mf{w}_\mr{p}}(k) + \bar{\mathcal{L}}_{\mf{w}_\mr{e}\mf{w}_\mr{p}}(k), \\
\bar{\Theta}_{\mf{w}_\mr{p}\mf{w}_\mr{p}} &= \bar{\Phi}_{\mf{w}_\mr{p}}(k)^\mr{T} \bar{V}_{\mf{yy}}(k+1) \bar{\Phi}_{\mf{w}_\mr{p}}(k) + \bar{\mathcal{L}}_{\mf{w}_\mr{p}\mf{w}_\mr{p}}(k), \\
\bar{\Theta}_{\mf{y}\mf{w}_\mr{e}} &= \bar{\Phi}_\mf{y}(k)^\mr{T} \bar{V}_{\mf{yy}}(k+1) \bar{\Phi}_{\mf{w}_\mr{e}}(k) + \bar{\mathcal{L}}_{\mf{y}\mf{w}_\mr{e}}(k), \\
\bar{\Theta}_{\mf{y}\mf{w}_\mr{p}} &= \bar{\Phi}_\mf{y}(k)^\mr{T} \bar{V}_{\mf{yy}}(k+1) \bar{\Phi}_{\mf{w}_\mr{p}}(k) + \bar{\mathcal{L}}_{\mf{y}\mf{w}_\mr{p}}(k), \\
\bar{\Theta}_{\mf{w}_\mr{e}\mf{y}} &= \bar{\Theta}_{\mf{y}\mf{w}_\mr{e}}^\mr{T}, \quad
\bar{\Theta}_{\mf{w}_\mr{p}\mf{y}} = \bar{\Theta}_{\mf{y}\mf{w}_\mr{p}}^\mr{T}, \quad
\bar{\Theta}_{\mf{w}_\mr{p}\mf{w}_\mr{e}} = \bar{\Theta}_{\mf{w}_\mr{e}\mf{w}_\mr{p}}^\mr{T}.
\end{align*}
This second step approximates how the same perturbations affect the current cost and the future cost-to-go. Minimizing this local model for the evader and maximizing it for the pursuer gives the local control update.

The perturbation constraint sets, following \eqref{eq:input_con}, reduce to
\begin{equation}
\begin{split}
\delta\mathcal{W}_\mr{e}&=\left\{\mf{u}^-_\mr{e}-\bar{\mf{u}}_\mr{e}(k)\le\delta\mf{w}_{\mr{e}(1:3)}\le\mf{u}^+_\mr{e}-\bar{\mf{u}}_\mr{e}(k),\right.\\
&\hspace{2.2cm}\left.\bm{\omega}^-_\mr{e}-\bar{\bm{\omega}}_\mr{e}(k)\le\delta\mf{w}_{\mr{e}(4:3+d)}\le\bm{\omega}^+_\mr{e}-\bar{\bm{\omega}}_\mr{e}(k)\right\}\\
&:=\left\{\mf{w}_\mr{e}^--\bar{\mf{w}}_\mr{e}(k)\le\delta\mf{w}_\mr{e}\le\mf{w}_\mr{e}^+-\bar{\mf{w}}_\mr{e}(k)\right\},
\\[0.5em]
\delta\mathcal{W}_\mr{p}&=\left\{\mf{u}^-_\mr{p}-\bar{\mf{u}}_\mr{p}(k)\le\delta\mf{w}_{\mr{p}(1:3)}\le\mf{u}^+_\mr{p}-\bar{\mf{u}}_\mr{p}(k),\right.\\
&\hspace{2.2cm}\left.\bm{\omega}^-_\mr{p}-\bar{\bm{\omega}}_\mr{p}(k)\le\delta\mf{w}_{\mr{p}(4:3+d)}\le\bm{\omega}^+_\mr{p}-\bar{\bm{\omega}}_\mr{p}(k)\right\}\\&:=\left\{\mf{w}_\mr{p}^--\bar{\mf{w}}_\mr{p}(k)\le\delta\mf{w}_\mr{p}\le\mf{w}_\mr{p}^+-\bar{\mf{w}}_\mr{p}(k)\right\}.
\end{split}
\end{equation}
Expanding the matrix product and dropping the time index $k$ for brevity, the second-order expansion becomes
\begin{equation}\label{eq:rhs}
\begin{split}
\min_{\delta\mathbf{w}_\mr{e}\in \delta \mathcal{W}_\mr{e}} \max_{\delta\mathbf{w}_\mr{p} \in \delta \mathcal{W}_\mr{p}} &\Big\{ \bar{\Theta} + \delta\mf{y}^\mr{T}\bar{\Theta}_{\mf{y}}+ \delta\mf{w}_\mr{e}^\mr{T}\bar{\Theta}_{\mf{w}_\mr{e}}+ \delta\mf{w}_\mr{p}^\mr{T}\bar{\Theta}_{\mf{w}_\mr{p}} + \frac{1}{2}\delta\mf{y}^\mr{T}\bar{\Theta}_{\mf{y}\mf{y}}\delta\mf{y} \\ 
&+ \frac{1}{2}\delta\mf{w}_\mr{e}^\mr{T}\bar{\Theta}_{\mf{w}_\mr{e}\mf{w}_\mr{e}}\delta\mf{w}_\mr{e} + \frac{1}{2}\delta\mf{w}_\mr{p}^\mr{T}\bar{\Theta}_{\mf{w}_\mr{p}\mf{w}_\mr{p}}\delta\mf{w}_\mr{p} + \delta\mf{w}_\mr{e}^\mr{T} \bar{\Theta}_{\mf{w}_\mr{e}\mf{w}_\mr{p}}\delta\mf{w}_\mr{p} + \delta\mf{w}_\mr{e}^\mr{T} \bar{\Theta}_{\mf{w}_\mr{e}\mf{y}}\,\delta\mf{y}+ \delta\mf{w}_\mr{p}^\mr{T} \bar{\Theta}_{\mf{w}_\mr{p}\mf{y}}\,\delta\mf{y} \Big\}.
\end{split}
\end{equation}
Since $\bar{\Theta}$ and $\delta\mf{y}^\mr{T} \bar{\Theta}_\mf{y} + \tfrac{1}{2}\delta\mf{y}^\mr{T} \bar{\Theta}_{\mf{y}\mf{y}}\delta\mf{y}$ do not depend on $(\delta\mf{w}_\mr{e}, \delta\mf{w}_\mr{p})$, they do not affect the argmin-argmax. Dropping them yields the equivalent quadratic program
\begin{equation}\label{eq:QP_joint}
\begin{split}
\min_{\delta\mf{w}_\mr{e}}\max_{\delta\mf{w}_\mr{p}} \quad
&\frac{1}{2}\delta\mf{w}_\mr{e}^\mr{T}
\bar{\Theta}_{\mf{w}_\mr{e}\mf{w}_\mr{e}}
\delta\mf{w}_\mr{e}
+\frac{1}{2}\delta\mf{w}_\mr{p}^\mr{T}
\bar{\Theta}_{\mf{w}_\mr{p}\mf{w}_\mr{p}}
\delta\mf{w}_\mr{p}
+\delta\mf{w}_\mr{e}^\mr{T}
\bar{\Theta}_{\mf{w}_\mr{e}\mf{w}_\mr{p}}
\delta\mf{w}_\mr{p}
\\
&+\left(
\bar{\Theta}_{\mf{w}_\mr{e}}
+\bar{\Theta}_{\mf{w}_\mr{e}\mf{y}}\delta\mf{y}
\right)^\mr{T}\delta\mf{w}_\mr{e}
+\left(
\bar{\Theta}_{\mf{w}_\mr{p}}
+\bar{\Theta}_{\mf{w}_\mr{p}\mf{y}}\delta\mf{y}
\right)^\mr{T}\delta\mf{w}_\mr{p}
\\
\mr{s.t.}\quad
&\mf{w}_\mr{e}^--\bar{\mf{w}}_\mr{e}
\le \delta\mf{w}_\mr{e}
\le \mf{w}_\mr{e}^+-\bar{\mf{w}}_\mr{e},
\\
&\mf{w}_\mr{p}^--\bar{\mf{w}}_\mr{p}
\le \delta\mf{w}_\mr{p}
\le \mf{w}_\mr{p}^+-\bar{\mf{w}}_\mr{p}.
\end{split}
\end{equation}

\subsubsection{Saddle-Point Solution}
We solve the constrained saddle-point problem \eqref{eq:QP_joint} at each node of the backward pass by adapting the primal active-set method of \cite{tassa2014control} to zero-sum games. The method fixes the active components of $\delta\mf{w}_\mr{e}$ and $\delta\mf{w}_\mr{p}$ at their bounds and computes the feedforward and feedback gains for the remaining free components. The active set is then updated using the Lagrange multipliers: a bound is released when its associated multiplier becomes negative, corresponding to a decrease in the objective for the evader or an increase in the objective for the pursuer. The resulting constrained feedforward and feedback gains are then substituted into the quadratic approximation to update the value function during the backward pass.

More specifically, les us form the Lagrangian
\begin{equation}\label{eq:lagrangian}
\begin{split}
\mathscr{L} &= \frac{1}{2}\delta\mf{w}_\mr{e}^\mr{T}\bar{\Theta}_{\mf{w}_\mr{e}\mf{w}_\mr{e}}\delta\mf{w}_\mr{e} + \frac{1}{2}\delta\mf{w}_\mr{p}^\mr{T}\bar{\Theta}_{\mf{w}_\mr{p}\mf{w}_\mr{p}}\delta\mf{w}_\mr{p} + \delta\mf{w}_\mr{e}^\mr{T} \bar{\Theta}_{\mf{w}_\mr{e}\mf{w}_\mr{p}}\delta\mf{w}_\mr{p} \\
&+ (\bar{\Theta}_{\mf{w}_\mr{e}} + \bar{\Theta}_{\mf{w}_\mr{e}\mf{y}}\delta\mf{y})^\mr{T}\delta\mf{w}_\mr{e} + (\bar{\Theta}_{\mf{w}_\mr{p}} + \bar{\Theta}_{\mf{w}_\mr{p}\mf{y}}\delta\mf{y})^\mr{T}\delta\mf{w}_\mr{p} \\ 
&+ \sum_{j=1}^{3+d} \lambda^+_{\mr{e}j}(\delta\mf{w}_{\mr{ej}}+\bar{\mf{w}}_{\mr{ej}}-\mf{w}^+_{\mr{ej}}) + \sum_{j=1}^{3+d} \lambda^-_{\mr{e}j}(\mf{w}^-_{\mr{ej}}-\delta\mf{w}_{\mr{ej}}-\bar{\mf{w}}_{\mr{ej}}) \\ 
&- \sum_{j=1}^{3+d} \lambda^+_{\mr{p}j}(\delta\mf{w}_{\mr{pj}}+\bar{\mf{w}}_{\mr{pj}}-\mf{w}^+_{\mr{pj}}) - \sum_{j=1}^{3+d} \lambda^-_{\mr{p}j}(\mf{w}^-_{\mr{pj}}-\delta\mf{w}_{\mr{pj}}-\bar{\mf{w}}_{\mr{pj}}),
\end{split}
\end{equation}
where $\lambda^+_{ij}, \lambda^-_{ij} \ge 0$ are the Lagrange multipliers for the upper and lower control bounds of player $i \in \{\mr{e}, \mr{p}\}$. Multipliers for the pursuer's constraints enter with a negative sign because the pursuer maximizes the objective. 

For each player $i\in\{\mr e,\mr p\}$, let $\mathcal A_i^+$ and $\mathcal A_i^-$ denote the index sets of the control components active at their upper and lower bounds, respectively, and let $\mathcal F_i$ denote the index set of free components. Under the concatenated control ordering, let $\mathcal A^+$, $\mathcal A^-$, and $\mathcal F$ denote the upper-active, lower-active, and free index sets of the joint control vector, respectively.  The joint perturbation under this ordering is partitioned as $\delta\mf w = [\delta\mf w_\mr e ~ \delta\mf w_\mr p]^\mr{T}$, $\delta\mf w_{\mathcal F} = [\delta\mf w_{\mr e}^{\mathcal F_\mr e} ~ \delta\mf w_{\mr p}^{\mathcal F_\mr p}]^\mr{T}$, and $\delta\mf w_{\mathcal A} = [\delta\mf w_{\mr e}^{\mathcal A_\mr e} ~ \delta\mf w_{\mr p}^{\mathcal A_\mr p}]^\mr{T}$. We define the concaneted active Lagrange multipliers $\Lambda_{\mathcal A}^{+} = [\Lambda_{\mathcal A_\mr e}^{+}~ -\Lambda_{\mathcal A_\mr p}^{+}]^\mr T$, and $\Lambda_{\mathcal A}^{-} =[\Lambda_{\mathcal A_\mr e}^{-}~-\Lambda_{\mathcal A_\mr p}^{-}]^\mr T$, while the multipliers associated with the free controls vanish, i.e., $\Lambda_{\mathcal F}^{+} = \Lambda_{\mathcal F}^{-}=\mf 0$. The first-order stationarity conditions for a saddle-point of \eqref{eq:lagrangian} can then be written as
\begin{align}\label{eq:partitioned_stationarity}
\begin{bmatrix} \bar{\Theta}_{\mathcal F}\\
\bar{\Theta}_{\mathcal A} + \Lambda^+_\mathcal{A} -\Lambda^-_\mathcal{A} \end{bmatrix} + \begin{bmatrix} \bar{\Theta}_{\mathcal F\mathcal F} & \bar{\Theta}_{\mathcal F\mathcal A}\\
\bar{\Theta}_{\mathcal A\mathcal F} & \bar{\Theta}_{\mathcal A\mathcal A} \end{bmatrix} \begin{bmatrix} \delta\mf w_{\mathcal F}^{\star}\\
\delta\mf w_{\mathcal A}^{\star}\end{bmatrix} + \begin{bmatrix} \bar{\Theta}_{\mathcal F\mf y}\\
\bar{\Theta}_{\mathcal A\mf y} \end{bmatrix} \delta\mf y = \mf 0.
\end{align}
where the vectors and matrices in \eqref{eq:partitioned_stationarity} are obtained from the original quantities by reordering their rows and columns so that the free control components precede the active control components.
Since the active controls are fixed at their corresponding bounds \cite{tassa2014control}, their feedforward and feedback gains satisfy 
\begin{align}\label{eq:active_gains}
\ell_{\mathcal A^+} &= \mf w_{\mathcal A^+}^{+} - \bar{\mf w}_{\mathcal A^+}, \\
\ell_{\mathcal A^-} &= \mf w_{\mathcal A^-}^{-} - \bar{\mf w}_{\mathcal A^-},\\
\mf K_{\mathcal A^+} &= \mf K_{\mathcal A^-} = \mf 0,
\end{align}
whereas the free feedforward and feedback gains are
\begin{align}\label{eq:gain_update}
\ell_{\mathcal F} &= -\bar{\Theta}_{\mathcal F\mathcal F}^{-1} \left( \bar{\Theta}_{\mathcal F} + \bar{\Theta}_{\mathcal F\mathcal A}\ell_{\mathcal A} \right),\\
\mf K_{\mathcal F} &= -\bar{\Theta}_{\mathcal F\mathcal F}^{-1} \bar{\Theta}_{\mathcal F\mf y}.
\end{align}
Reordering the constrained gains according to the original control ordering gives the local affine policy $\delta\mf w^\star=\ell+\mf K\,\delta\mf y$.

To verify that the assumed active set satisfies the KKT conditions, the corresponding Lagrange multipliers are evaluated from the active stationarity conditions. The constant terms in the active stationarity conditions satisfy
\begin{align}
\bar{\Theta}_{\mathcal A}
+\bar{\Theta}_{\mathcal A\mathcal F}\ell_{\mathcal F}
+\bar{\Theta}_{\mathcal A\mathcal A}\ell_{\mathcal A}
+\Lambda_{\mathcal A}^{+}
-\Lambda_{\mathcal A}^{-}
&=
\mf 0.
\end{align}
Complementary slackness requires
$\Lambda_{\mathcal A^+}^{-}=\mf 0$ and
$\Lambda_{\mathcal A^-}^{+}=\mf 0$. The active multipliers are
therefore
\begin{align}\label{eq:multiplier_eval}
j\in\mathcal A_\mr e^+:\quad
\lambda_{\mr e j}^+
&=
-\left[
\bar{\Theta}_{\mathcal A}
+\bar{\Theta}_{\mathcal A\mathcal F}\ell_{\mathcal F}
+\bar{\Theta}_{\mathcal A\mathcal A}\ell_{\mathcal A}
\right]_j,
&
j\in\mathcal A_\mr e^-:\quad
\lambda_{\mr e j}^-
&=
\left[
\bar{\Theta}_{\mathcal A}
+\bar{\Theta}_{\mathcal A\mathcal F}\ell_{\mathcal F}
+\bar{\Theta}_{\mathcal A\mathcal A}\ell_{\mathcal A}
\right]_j,
\\
j\in\mathcal A_\mr p^+:\quad
\lambda_{\mr p j}^+
&=
\left[
\bar{\Theta}_{\mathcal A}
+\bar{\Theta}_{\mathcal A\mathcal F}\ell_{\mathcal F}
+\bar{\Theta}_{\mathcal A\mathcal A}\ell_{\mathcal A}
\right]_j,
&
j\in\mathcal A_\mr p^-:\quad
\lambda_{\mr p j}^-
&=
-\left[
\bar{\Theta}_{\mathcal A}
+\bar{\Theta}_{\mathcal A\mathcal F}\ell_{\mathcal F}
+\bar{\Theta}_{\mathcal A\mathcal A}\ell_{\mathcal A}
\right]_j.
\end{align}
Dual feasibility requires $\lambda_{ij}^{\pm}\geq0$. If an active
constraint yields a negative multiplier, the corresponding control is
released into $\mathcal F$, and the active-set iteration repeats until
the KKT conditions are satisfied.

\subsubsection{Backward Recursions}
Having characterized the optimal control perturbations, we use them to obtain a local approximation of the value function backward in time. The core of this process is the backward pass: at each node $k$, we use the saddle-point solution of the local quadratic game to update the quadratic approximation of $V(\mf{y}(k), k)$, which then serves as the boundary condition for the game at node $k-1$.
To this end, a second-order approximation of the left-hand side of \eqref{eq:HJI} yields
\begin{equation}\label{eq:lhs}
\bar{V}(\mf{y}(k)) \approx \bar{V}(k) + \bar{V}_\mf{y}(k)^\mr{T} \delta\mf{y}(k) + \tfrac{1}{2}\delta\mf{y}(k)^\mr{T} \bar{V}_{\mf{yy}}(k) \delta\mf{y}(k).
\end{equation}
Enforcing \eqref{eq:HJI} amounts to equating this left-hand side approximation to the right-hand side approximation \eqref{eq:rhs} under the local affine policies derived above. Matching terms of equal order in $\delta\mf{y}(k)$, we obtain the backward recursions for $\bar{V}, \bar{V}_\mf{y}, \bar{V}_{\mf{yy}}$:
\begin{equation}\label{eq:backward}
\begin{split}
\bar{V}(k) &= \bar{\Theta} + \ell_{\mf{w}_\mr{e}}^\mr{T} \bar{\Theta}_{\mf{w}_\mr{e}} + \ell_{\mf{w}_\mr{p}}^\mr{T} \bar{\Theta}_{\mf{w}_\mr{p}} + \tfrac{1}{2}\Big( \ell_{\mf{w}_\mr{e}}^\mr{T} \bar{\Theta}_{\mf{w}_\mr{e}\mf{w}_\mr{e}} \ell_{\mf{w}_\mr{e}} + \ell_{\mf{w}_\mr{p}}^\mr{T} \bar{\Theta}_{\mf{w}_\mr{p}\mf{w}_\mr{p}} \ell_{\mf{w}_\mr{p}} \\
& \quad + \ell_{\mf{w}_\mr{e}}^\mr{T} \bar{\Theta}_{\mf{w}_\mr{e}\mf{w}_\mr{p}} \ell_{\mf{w}_\mr{p}} + \ell_{\mf{w}_\mr{p}}^\mr{T} \bar{\Theta}_{\mf{w}_\mr{p}\mf{w}_\mr{e}} \ell_{\mf{w}_\mr{e}} \Big), \\
\bar{V}_\mf{y}(k) &= \bar{\Theta}_\mf{y} + \mf{K}_{\mf{w}_\mr{e}}^\mr{T} \bar{\Theta}_{\mf{w}_\mr{e}} + \mf{K}_{\mf{w}_\mr{p}}^\mr{T} \bar{\Theta}_{\mf{w}_\mr{p}} + \bar{\Theta}_{\mf{y}\mf{w}_\mr{e}} \ell_{\mf{w}_\mr{e}} + \bar{\Theta}_{\mf{y}\mf{w}_\mr{p}} \ell_{\mf{w}_\mr{p}} + \mf{K}_{\mf{w}_\mr{e}}^\mr{T} \bar{\Theta}_{\mf{w}_\mr{e}\mf{w}_\mr{e}} \ell_{\mf{w}_\mr{e}} \\
& \quad + \mf{K}_{\mf{w}_\mr{p}}^\mr{T} \bar{\Theta}_{\mf{w}_\mr{p}\mf{w}_\mr{p}} \ell_{\mf{w}_\mr{p}} + \mf{K}_{\mf{w}_\mr{e}}^\mr{T} \bar{\Theta}_{\mf{w}_\mr{e}\mf{w}_\mr{p}} \ell_{\mf{w}_\mr{p}} + \mf{K}_{\mf{w}_\mr{p}}^\mr{T} \bar{\Theta}_{\mf{w}_\mr{p}\mf{w}_\mr{e}} \ell_{\mf{w}_\mr{e}}, \\
\bar{V}_{\mf{yy}}(k) &= \bar{\Theta}_{\mf{yy}} + \mf{K}_{\mf{w}_\mr{e}}^\mr{T} \bar{\Theta}_{\mf{w}_\mr{e}\mf{y}} + \bar{\Theta}_{\mf{y}\mf{w}_\mr{e}} \mf{K}_{\mf{w}_\mr{e}} + \mf{K}_{\mf{w}_\mr{p}}^\mr{T} \bar{\Theta}_{\mf{w}_\mr{p}\mf{y}} + \bar{\Theta}_{\mf{y}\mf{w}_\mr{p}} \mf{K}_{\mf{w}_\mr{p}} + \mf{K}_{\mf{w}_\mr{e}}^\mr{T} \bar{\Theta}_{\mf{w}_\mr{e}\mf{w}_\mr{e}} \mf{K}_{\mf{w}_\mr{e}} \\
& \quad + \mf{K}_{\mf{w}_\mr{p}}^\mr{T} \bar{\Theta}_{\mf{w}_\mr{p}\mf{w}_\mr{p}} \mf{K}_{\mf{w}_\mr{p}} + \mf{K}_{\mf{w}_\mr{e}}^\mr{T} \bar{\Theta}_{\mf{w}_\mr{e}\mf{w}_\mr{p}} \mf{K}_{\mf{w}_\mr{p}} + \mf{K}_{\mf{w}_\mr{p}}^\mr{T} \bar{\Theta}_{\mf{w}_\mr{p}\mf{w}_\mr{e}} \mf{K}_{\mf{w}_\mr{e}}.
\end{split}
\end{equation}
Finally, under a second-order approximation of the boundary condition in \eqref{eq:HJI}, these are subject to 
\begin{equation}\label{eq:boundary}
\bar{V}(N)=\phi(\kappa(\bar{\mf{y}}(N)), t(N)),\quad
\bar{V}_{\mf{y}}(N)=\phi_{\mf{y}}(\kappa(\bar{\mf{y}}(N)), t(N)),\quad
\bar{V}_{\mf{yy}}(N)=\phi_{\mf{yy}}(\kappa(\bar{\mf{y}}(N)), t(N)).
\end{equation}
DDP then iteratively solves the forward perturbation equation \eqref{eq:forward} and the backward equations \eqref{eq:backward} until convergence.

\subsection{Regularization and Line Search}
In discrete-time implementations of DDP, it is common for the Hessian $\bar{\mr{\Theta}}_{\mf{w}_\mr{e}\mf{w}_\mr{e}}$, and by extension the Hessian $\bar{\mr{\Theta}}_{\mf{w}_\mr{p}\mf{w}_\mr{p}}$, to become indefinite during the backward pass. This loss of definiteness can cause divergence of the algorithm and therefore motivates the use of regularization to enforce strict definiteness \cite{tassa2014control}. To this end, at each backward pass of DDP, we add a sufficiently large regularization parameter $\lambda \ge 0$ to substitute $\bar{\mr{\Theta}}_{\mf{w}_\mr{e}\mf{w}_\mr{e}}$ and $\bar{\mr{\Theta}}_{\mf{w}_\mr{p}\mf{w}_\mr{p}}$ with $\bar{\mr{\Theta}}_{\mf{w}_\mr{e}\mf{w}_\mr{e}}+\lambda I_{3+d}$ and $\bar{\mr{\Theta}}_{\mf{w}_\mr{p}\mf{w}_\mr{p}}-\lambda I_{3+d}$, respectively. This substitution ensures the backward pass remains well-posed and the control gains are well-defined. Importantly, this regularization is restricted to the inversion of the Hessians; the rest of the matrices are preserved when updating the value function approximations $\bar{V}_\mf{y}$ and $\bar{V}_{\mf{y}\mf{y}}$ to prevent distortion of the true cost-to-go.

Furthermore, because the backward pass relies on a local quadratic approximation, using large step sizes in the DDP update can also lead to divergence. Therefore, to identify a proper step size, we employ a backtracking line search on the feedforward term coupled with a trust-region update during the forward pass\cite{tassa2012synthesis}\cite{nocedal2006numerical}. 

More specifically, let $\Delta V_1$ and $\Delta V_2$ denote the first- and second-order expected cost reductions computed during the backward pass over the $N$ discrete time steps:
\begin{equation}
\begin{split}\label{eq:expected_cost}
    \Delta V_1 &= \sum_{k=0}^{N-1} \bar{\Theta}_{\mf{w}_\mr{e}}^\mr{T}(k) \ell_{\mf{w}_\mr{e}}(k) + \bar{\Theta}_{\mf{w}_\mr{p}}^\mr{T}(k) \ell_{\mf{w}_\mr{p}}(k), \\
    \Delta V_2 &= \sum_{k=0}^{N-1} \ell_{\mf{w}_\mr{e}}^\mr{T}(k) \bar{\Theta}_{\mf{w}_\mr{e}\mf{w}_\mr{e}} \ell_{\mf{w}_\mr{e}}(k) + \ell_{\mf{w}_\mr{p}}^\mr{T}(k) \bar{\Theta}_{\mf{w}_\mr{p}\mf{w}_\mr{p}} \ell_{\mf{w}_\mr{p}}(k) + 2 \ell_{\mf{w}_\mr{e}}^\mr{T}(k) \bar{\Theta}_{\mf{w}_\mr{e}\mf{w}_\mr{p}} \ell_{\mf{w}_\mr{p}}(k).
\end{split}
\end{equation}
During the forward pass, the feedforward control updates are scaled by a step size $\alpha \in (0, 1]$. The expected change in cost for a given $\alpha$ is given by the quadratic model:
\begin{align}
    \Delta J_{e}(\alpha) = \alpha \Delta V_1 + \frac{\alpha^2}{2} \Delta V_2.
\end{align}
For each candidate trajectory, we evaluate the actual non-linear change in the total cost, and compute the trust-region ratio:
\begin{align}
    \rho = \frac{\sum_{k=0}^{N-1} J(\{\mathbf{w}_{\mr{e}}(k)\},\{\mathbf{w}_{\mr{p}}(k)\})-J(\{\hat{\mathbf{w}}_{\mr{e}}(k)\},\{\hat{\mathbf{w}}_{\mr{p}}(k)\})}{\Delta J_{e}(\alpha)}
\end{align}
where $\hat{\mathbf{w}}_\mr{e}(k) = \mathbf{w}_\mr{e}(k)+\alpha \ell_{\mf{w}_\mr{e}}(k)+\mf{K}_{\mf{w}_\mr{e}}(k)\delta\mathbf{y}(k)$ and $\hat{\mathbf{w}}_\mr{p}(k) = \mathbf{w}_\mr{p}(k)+\alpha \ell_{\mf{w}_\mr{p}}(k)+\mf{K}_{\mf{w}_\mr{p}}(k)\delta\mathbf{y}(k)$. The parameter $\rho$ serves as a direct measure of the local model's fidelity and governs the adaptation of the regularization parameter $\lambda$. If $\rho \approx 1$, the step is accepted and $\lambda$ is decreased to expand the trust region; if $\rho$ is poor but positive ($\rho \le \varepsilon_1$, $\varepsilon_1\in[0,1)$), the step is accepted but $\lambda$ is increased to shrink the trust region.  Conversely, if the actual cost reduction is excessively large compared to the model's prediction ($\rho \ge \varepsilon_2$, $\varepsilon_2>1$), the local approximation is deemed fundamentally untrustworthy. 

Note that, in single-player optimization, $\rho \gg 1$ is typically benign and the step would be accepted. Here, however, $J_\mr{e}(\alpha)$ is computed assuming both players respond optimally within their local quadratic models. A large $\rho$, therefore, cannot be attributed solely to model conservatism; it may equally reflect a failure of the opponent's quadratic model to predict their true best response, producing an improvement that exploits a modeling artifact rather than a genuine game-theoretic improvement. Since such an improvement is unlikely to persist once the opponent adapts, the step is rejected. In such cases, or if no step size $\alpha$ yields a valid improvement ($\rho > 0$), the step is rejected entirely. The backward pass is then re-computed with a larger $\lambda$ to enforce a more conservative step.

The complete game-theoretic DDP algorithm, incorporating these details, is summarized in Algorithm \ref{al:ddp}.

\begin{algorithm}
\caption{Discrete Game-Theoretic DDP}
\hspace*{\algorithmicindent} \textbf{Input}: Initial state trajectory $\bar{\mf{y}}$, initial controls $\{\bar{\mf{w}}_\mr{e}, \bar{\mf{w}}_\mr{p}\}$, initial $\lambda_\text{reg} > 0$, tolerance $\varepsilon>0$.\\
\hspace*{\algorithmicindent} \textbf{Output}: Evader and pursuer policies $\mf{w}_\mr{e}^\star, \mf{w}_\mr{p}^\star$, joint trajectory $\mf{y}^\star$.
\begin{algorithmic}[1]
\Procedure{}{}
\State
$\delta\mathbf{w}_\mr{e}\leftarrow\infty$, $\delta\mathbf{w}_\mr{p}\leftarrow\infty$
\While{$\|\delta\mathbf{w}_\mr{e}\| + \|\delta\mathbf{w}_\mr{p}\| > \varepsilon$}
    \State Compute linearizations $\bar{\Phi}_{\mf{y}},\bar{\Phi}_{\mf{w}_{\mr{e}}},\bar{\Phi}_{\mf{w}_{\mr{p}}}$, cost derivatives along $\{\bar{\mf{y}}, \bar{\mathbf{w}}_\mr{e}, \bar{\mathbf{w}}_\mr{p}\}$, and current cost $J$.
    \State \textbf{Backward Pass:} Initialize $\bar{V}_{\mf{y}}, \bar{V}_{\mf{y}\mf{y}}$ at step $N$ via \eqref{eq:boundary}.
    \For{$k = N-1$ \textbf{down to} $0$}
        \State Construct $\bar{\Theta}_{\mf{w}_\mr{e}\mf{w}_\mr{e},\text{reg}} \leftarrow \bar{\Theta}_{\mf{w}_\mr{e}\mf{w}_\mr{e}} + \max(\lambda_\mr{reg},  10^{-7}) I_{3+d}$ and $\bar{\Theta}_{\mf{w}_\mr{p}\mf{w}_\mr{p},\text{reg}} \leftarrow \bar{\Theta}_{\mf{w}_\mr{p}\mf{w}_\mr{p}} -  \max(\lambda_\mr{reg},  10^{-7}) I_{3+d}$.
        \If{$\bar{\Theta}_{\mf{w}_\mr{e}\mf{w}_\mr{e},\text{reg}} \not\succ 0$ \textbf{or} $-\bar{\Theta}_{\mf{w}_\mr{p}\mf{w}_\mr{p},\text{reg}} \not\succ 0$}
            \State $\lambda_\text{reg} \leftarrow  10\lambda_\text{reg}$ and \textbf{restart} Backward Pass.
        \EndIf
        \State Solve for constrained steps $\ell_{\mf{w}_\mr{e}}(k), \ell_{\mf{w}_\mr{p}}(k)$ and feedback gains $\mf{K}_{\mf{w}_\mr{e}}(k), \mf{K}_{\mf{w}_\mr{p}}(k)$ from \eqref{eq:gain_update}, update $\bar{V}_\mf{y}, \bar{V}_{\mf{y}\mf{y}}$ using \eqref{eq:backward}, and accumulate $\Delta V_1, \Delta V_2$ in \eqref{eq:expected_cost}.
    \EndFor
    \State \textbf{Forward Pass \& Trust-Region Line Search:} Set $\text{valid} \leftarrow \text{false}$.
    \For{$\alpha \in \{\alpha_1,\alpha_2, \dots\}$}
        \State Generate new trajectory $\hat{\mf{y}}$ under $\hat{\mathbf{w}}_\mr{e}(k) = \bar{\mathbf{w}}_\mr{e}(k) + \alpha \ell_{\mf{w}_\mr{e}}(k) + \mf{K}_{\mf{w}_\mr{e}}(k)\,\delta\mathbf{y}(k)$ and $\hat{\mathbf{w}}_\mr{p}(k) = \bar{\mathbf{w}}_\mr{p}(k) + \alpha \ell_{\mf{w}_\mr{p}}(k) + \mf{K}_{\mf{w}_\mr{p}}(k)\,\delta\mathbf{y}(k)$.
        \State Compute candidate cost $\hat{J}$ and trust-region ratio $\rho \leftarrow (J - \hat{J}) / (\alpha \Delta V_1 + \tfrac{\alpha^2}{2} \Delta V_2)$.
        \If{$0 < \rho < \varepsilon_2$}
            \State $\text{valid} \leftarrow \text{true}$
            \State Update $\lambda_\text{reg} \leftarrow \lambda_\text{reg}/3$ (if $\rho \approx 1$), $\lambda_\text{reg} \leftarrow \lambda_\text{reg}/1.5$ (if $\rho \geq \varepsilon_1$), or else $\lambda_\text{reg} \leftarrow 2\lambda_\text{reg}$.
            \State \textbf{break} \Comment{Exit line search}
        \EndIf
    \EndFor
    \If{\text{valid}}
        \State Update $\delta\mathbf{w}_\mr{e} \leftarrow\hat{\mathbf{w}}_\mr{e}-\bar{\mathbf{w}}_\mr{e}$ and $\delta\mathbf{w}_\mr{p} \leftarrow\hat{\mathbf{w}}_\mr{p}-\bar{\mathbf{w}}_\mr{p}$.
        \State Update $\bar{\mf{y}} \leftarrow \hat{\mf{y}}$, $\bar{\mf{w}}_\mr{e} \leftarrow \hat{\mf{w}}_\mr{e}$, $\bar{\mf{w}}_\mr{p} \leftarrow \hat{\mf{w}}_\mr{p}$.
    \Else
        \State $\lambda_\text{reg} \leftarrow 10\lambda_\text{reg}$ and \textbf{restart} Backward Pass
    \EndIf
\EndWhile
\State \textbf{return} $\mf{y}^\star \leftarrow \bar{\mf{y}}$, $\mf{w}_\mr{e}^\star \leftarrow \bar{\mf{w}}_\mr{e}$, $\mf{w}_\mr{p}^\star \leftarrow \bar{\mf{w}}_\mr{p}$
\EndProcedure
\end{algorithmic}\label{al:ddp}
\end{algorithm}

\section{Numerical Results}
\begin{figure}[t]
    \centering
    \includegraphics{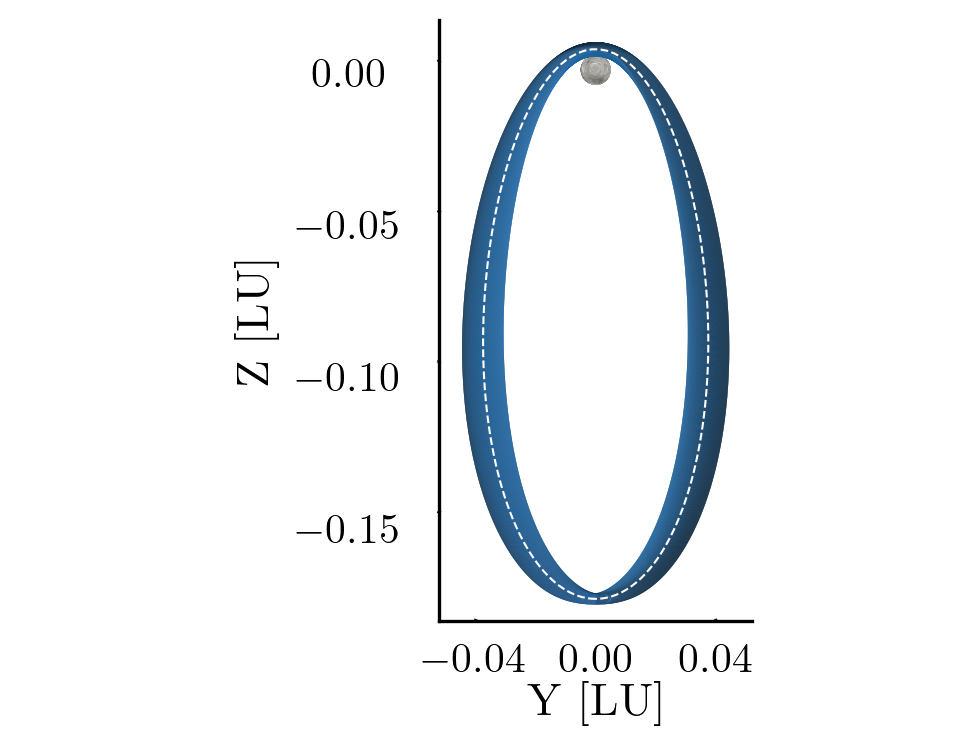}
    \caption{$L_2$ southern NRHO periodic orbit and quasi-periodic orbit. The 3D plot illustrates the invariant torus. The white dashed line corresponds to the PO from which the QPO family extends. The blue color denotes the QPO generated.}
    \label{fig:NRHO}
\end{figure}

We now present numerical results for pursuit-evasion scenarios over periodic and quasi-periodic orbits. We first compare the discrete formulation proposed here with the continuous formulation developed in~\cite{fotiadis2026adversarial} on the periodic NRHO shown in Fig.~\ref{fig:NRHO}. The surrounding quasi-NRHO tori are also included because they are generated from this periodic orbit and are used in the later quasi-periodic analysis. We then study how running the game on a quasi-periodic orbit changes the resulting maneuvers and evaluate the added benefit of a second phase-control input. Finally, we compare this quasi-NRHO with a quasi-halo orbit from the family shown in Fig.~\ref{fig:QPO_family}. These two families are selected because they have different lunar-approach geometries and stability properties, which allows us to study how the reference orbit affects the pursuit-evasion outcome.
\subsection{Continuous vs Discrete DDP}
\label{sec:CT_vs_DT}
We consider an evader-pursuer interaction on NRHO shown in Fig.~\ref{fig:NRHO}. The orbit has a period of $T=6.5~\mathrm{days}$ ($1.466695~[\mathrm{ND}]$). The evader seeks to increase the separation beyond $600~\mathrm{km}$, whereas the pursuer seeks to minimize it. Both spacecraft have mass $m_\mathrm{e}=m_\mathrm{p}=1000~\mathrm{kg}$ and remain near the reference orbit. Initially, the pursuer starts at apolune, and the evader starts approximately $170$ km ahead along the orbit.

We compare the proposed discrete-time differential dynamic programming method (DT-DDP) with a continuous-time DDP formulation (CT-DDP) from \cite{fotiadis2026adversarial}. We select
\begin{align*}
&R_\mathrm{e}=5\cdot10^{-3}\cdot I_3,\quad R_\mathrm{p}=10^{-2}I_3,\\
&A_\mathrm{e}=5\cdot10^{-3},\quad A_\mathrm{p}=10^{-2},\\
&Q_{\mathrm{e}}=F_{\mathrm{e}}=Q_{\mathrm{p}}=F_{\mathrm{p}}=5I_6,\\
& w=2000,\quad p=2.1,
\end{align*}
which makes the evader twice as maneuverable as the pursuer. We set the escape threshold to $D_0=660~\mathrm{km}$, which exceeds the mission requirement by $10\%$ to compensate for the vanishing behavior of~\eqref{eq:evadecost} near $D_0$. 

The Cartesian velocity recovered from the KS variables using Eq.~\eqref{eq:conv_KS_cart} contains a factor $1/r_\mr{i}$, which can lead to poor numerical conditioning near the Moon. To reduce this effect while keeping the tracking cost in Cartesian coordinates, we scale the velocity weights using the corresponding player position:
\begin{align*}
F_\mr{i}(r_\mr{i})=Q_\mr{i}(r_\mr{i})=\begin{bmatrix}5I_3&0_3\\0_3&1000 r_\mr{i}^{\,2}I_3\end{bmatrix},\qquad i\in\{\mathrm e,\mathrm p\}.
\end{align*}
The factor $r_\mr{i}^{\,2}$ compensates for the $1/r_\mr{i}$ dependence of the Cartesian velocity conversion and prevents the velocity error from dominating the tracking cost near close lunar approaches. Since these close approaches occupy only a small portion of the trajectory, this scaling primarily affects the cost locally near the Moon while leaving the weighting elsewhere largely unchanged. The additional factor $200$ balances the position and velocity contributions over the NRHO, where the average lunar distance is approximately $0.04$~LU.

Although the tracking cost remains well conditioned under this weighting, the dynamics themselves become highly sensitive near perilune, making the game increasingly difficult to solve numerically. The proposed multi-agent time regularization addresses this issue by introducing a shared fictitious time that automatically increases temporal resolution when either spacecraft approaches the Moon. DT-DDP then operates directly on the resulting regularized discretization during both the forward and backward sweeps, whereas CT-DDP computes its backward sweep by integrating continuous-time value-function equations. Consequently, the discrete-time formulation is expected to provide a numerical advantage during close lunar encounters.

\begin{figure}
    \centering
    \begin{subfigure}{\columnwidth}
         \centering
         \includegraphics{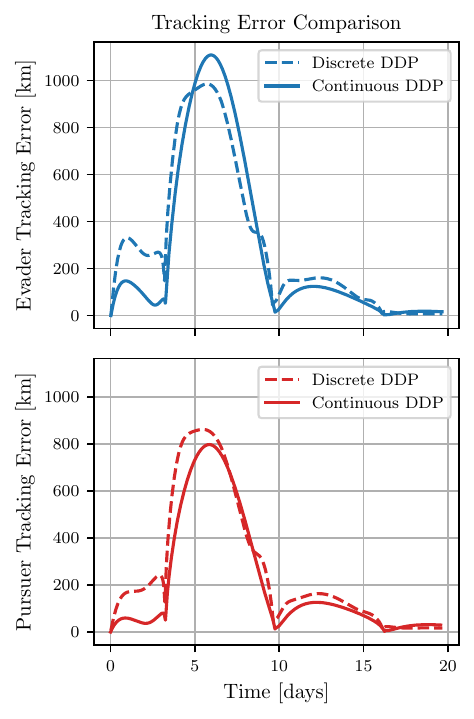}
         \caption{Evader and pursuer tracking errors. Both formulations keep the tracking errors bounded and produce similar overall trends, with differences concentrated near the apolune.}
         \label{fig:DT_vs_CT_tracking}
     \end{subfigure}
     \begin{subfigure}{\columnwidth}
         \centering
         \includegraphics{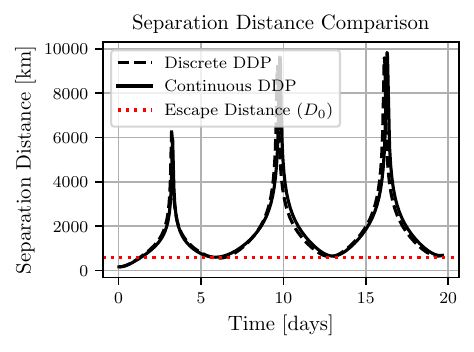}
         \caption{Separation distance between the agents. Both formulations exceed the prescribed escape distance and produce similar separation peaks over the three orbital revolutions.}
         \label{fig:DT_vs_CT_separation}
     \end{subfigure}
     \caption{Comparison between the continuous-time and discrete-time DDP formulations. Both methods keep the agents near the reference orbit and achieve separation above the escape distance, although they produce different tracking profiles because of their different discretizations and tracking costs.}
\end{figure}
Figures \ref{fig:DT_vs_CT_tracking}-\ref{fig:DT_vs_CT_separation} compare the tracking and the separation results. Both methods follow the reference orbit and are able to escape, but DT-DDP produces different trajectory profiles due to the discretization and the tracking weight difference. Because the tracking matrix is scaled by the agent position, the optimization metric remains fundamentally different from the physical Cartesian tracking used in CT-DDP, resulting in distinct but similar trajectory profiles. Despite these differences, the discrete-time formulation provided a substantial computational advantage. DT-DDP completed the simulation in $10.01$~s, whereas CT-DDP required $302.73$~s, corresponding to an observed speedup of approximately $30.2\times$. 

\subsection{Regularized vs Uniform Discretization}
We next look at the benefit of the proposed multi-agent time regularization by comparing it against a uniform, non-regularized time discretization for the same pursuit-evasion game. Both schemes solve the same NRHO scenario of Section~\ref{sec:CT_vs_DT}, with the same number of points. To show the effect of close lunar approaches, we vary the periapsis distance of the reference orbit and record the wall-clock run time required for DT-DDP to converge under each discretization scheme. 

Table~\ref{tab:comp_disc} summarizes the results. When the periapsis distance is large, the two schemes perform comparably, since the dynamics remain well-conditioned throughout the game and the adaptive discretization approaches a nearly uniform physical-time distribution. As the periapsis distance decreases, however, the uniform discretization degrades sharply: its fixed time step under-resolves the rapidly varying dynamics near the perilune, forcing additional DDP iterations to compensate. In contrast, the regularized scheme automatically concentrates nodes near the close approach, maintaining convergence and a run time that grows only modestly as the periapsis distance shrinks. These results confirm that the multi-agent time regularization yields an important computational advantage for scenarios with close encounters with a primary.

\begin{table}[h]
    \centering
    \caption{Multi-Agent Time Regularization vs Uniform Time Discretization as a function of the Periapsis Distance}
    \label{tab:comp_disc}
    \begin{tabular}{l c c c c}
        \hline
        Periapsis Distance & Regularized Time & Regularized Iters & Non-Regularized Time & Non-Regularized Iters \\
        \hline
        $1894.56$ km & $12.83$ s & $132$ & $54.64$ s & $532$ \\
        $2782.17$ km & $8.26$ s & $85$ & $9.26$ s & $101$ \\
        $25715.76$ km & $4.87$ s & $47$ & $5.36$ s & $55$ \\
        \hline
    \end{tabular}
\end{table}

\subsection{Pursuit-Evasion on a QPO}
We study how a quasi-periodic reference and the added transverse phase control affect the pursuit-evasion outcome, we evaluate the differential game on the quasi-NRHO family shown in Fig.~\ref{fig:NRHO}. The reference trajectory is defined by the two-dimensional Fourier representation $\hat{K}(\theta_1, \theta_2)$ constructed in Section IV.B.2. The spacecraft initialize their states on the invariant torus with an initial along-track separation of approximately $166$ km. The spacecraft masses, thrust constraints, and cost penalties match the periodic orbit baseline. The cost parameters are kept the same as before, aside from the tracking and control weights and the addition of the transverse direction 
\begin{align*}
    &F_\mr{i}(r_\mr{i})=Q_\mr{i}(r_\mr{i})=50\begin{bmatrix}I_3&0_3\\0_3& r_\mr{i}^{\,2}I_3\end{bmatrix},\qquad i\in\{\mathrm e,\mathrm p\},\\
    &R_\mr{e} = 5\cdot10^{-4}I_3, \quad R_\mr{p}=10^{-3}I_3,\\
    &A_e = \begin{bmatrix}
        5\cdot10^{-3} & 0 \\ 0 & 5\cdot10^{-6}
    \end{bmatrix},
    \quad A_\mathrm{p} = \begin{bmatrix}
        10^{-2} & 0 \\ 0 & \cdot10^{-5}
    \end{bmatrix}.
\end{align*}
We also add the following control input constraints:
\begin{align*}
\mf{u}^-_\mr{e} = \begin{bmatrix}
    -0.5 \\ - 0.5 \\ -0.5
\end{bmatrix},~
\mf{u}^+_\mr{e} = \begin{bmatrix}
    0.5 \\  0.5 \\ 0.5
\end{bmatrix},~
\mf{u}^-_\mr{p} = \begin{bmatrix}
    -0.2 \\  -0.2 \\ -0.2
\end{bmatrix},~
\mf{u}^+_\mr{p} = \begin{bmatrix}
    0.2 \\  0.2 \\ 0.2
\end{bmatrix}.
\end{align*}

In this scenario, both agents utilize full two-dimensional phase control $\omega_\mr{i} = [\omega_{\mr{i},1}, \omega_{\mr{i},2}]^\mr{T}$. Each player varies its longitudinal phase rate  $\omega_{\mr{i},1}$ to accelerate or decelerate along the track. Concurrently, they adjust the transverse phase rate $\omega_{\mr{i},2}$ to shift its position laterally across the torus surface. This allows both spacecraft to steer their reference signals dynamically to any point on the two-dimensional manifold rather than remaining restricted to a one-dimensional curve.  

\begin{figure}
    \centering
    \begin{subfigure}{\columnwidth}
         \centering
         \includegraphics{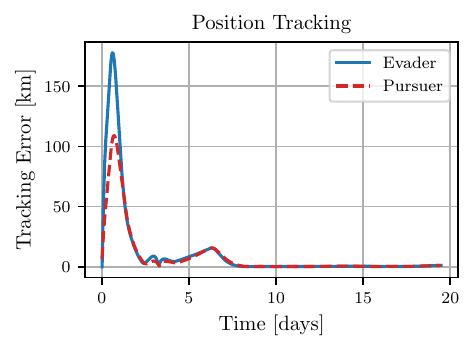}
         \caption{Position tracking errors over the quasi-NRHO. The initial maneuver produces a short increase in tracking error, after which both agents return close to the reference.}
         \label{fig:QPO_tracking}
     \end{subfigure}
     \begin{subfigure}{\columnwidth}
         \centering
        \includegraphics{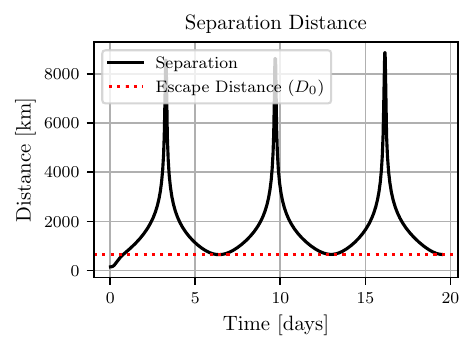}
         \caption{Separation distance over the quasi-NRHO. The evader quickly exceeds the escape distance and maintains the required separation throughout the encounter.}
         \label{fig:QPO_separation}
     \end{subfigure}
     \begin{subfigure}{\columnwidth}
        \centering
        \includegraphics{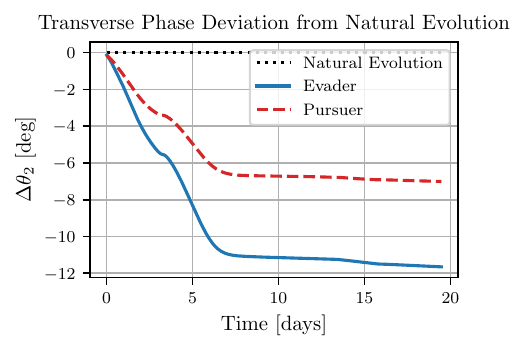}
         \caption{Transverse phase deviations from natural evolution. Both agents adjust their transverse phases, with the evader using a larger deviation to transition to a different torus fiber.}
         \label{fig:QPO_transverse}
     \end{subfigure}
     \caption{Pursuit-evasion over a quasi-NRHO. The agents use two-dimensional phase control to increase their separation while remaining close to the quasi-periodic reference. The transverse phase adjustment allows the evader to move to a different torus fiber rather than relying only on motion away from the reference.}
\end{figure}

The game results show fast separation similar to the PO case, but with a smaller separation from the reference trajectory as shown in Fig.~\ref{fig:QPO_tracking}. As demonstrated in Fig.~\ref{fig:QPO_transverse}, the evader achieves this by transitioning to a different torus fiber (a torus fiber is the closed curve traced by the natural flow from a fixed transverse phase). Because it moves along the torus surface, the evader does not need to deviate as far from the reference as in the PO case. This strategy leads to faster convergence to an escape, i.e., the evader quickly returns to the reference trajectory, while maintaining the required separation distance, as shown in Fig.~\ref{fig:QPO_separation}.

\subsection{Quasi Halo vs Quasi NRHO}
To highlight how lunar-approach geometry and orbital instability affect the pursuit-evasion outcome, we compare the game on the quasi-halo and quasi-NRHO families. Although both scenarios provide the agents with the same thrust and two-dimensional phase-control inputs, they evolve in different gravitational environments and have different local stability properties. The maximum eigenvalue magnitude of the monodromy matrix is approximately $3.7$ for the underlying halo orbit and $1.9$ for the NRHO, indicating stronger perturbation growth over one period near the halo orbit. In contrast, the quasi-NRHO passes much closer to the Moon and is therefore subject to stronger gravitational gradients during perilune. To make these differences more visible, we reduce the tracking weights and the control limits while increasing the control penalties
\begin{align*}
    &F_\mr{i}(r_\mr{i})=Q_\mr{i}(r_\mr{i})=20\begin{bmatrix}I_3&0_3\\0_3& r_\mr{i}^{\,2}I_3\end{bmatrix},\qquad i\in\{\mathrm e,\mathrm p\},\\
    &R_\mr{e} = 5\cdot10^{-3}I_3, \quad R_\mr{p}=10^{-2}I_3,\\
    &\mf{u}^-_\mr{e} = \begin{bmatrix}
    -0.05 \\ - 0.05 \\ -0.05
    \end{bmatrix},~
    \mf{u}^+_\mr{e} = \begin{bmatrix}
        0.05 \\  0.05 \\ 0.05
    \end{bmatrix},~
    \mf{u}^-_\mr{p} = \begin{bmatrix}
        -0.02 \\  -0.02 \\ -0.02
    \end{bmatrix},~
    \mf{u}^+_\mr{p} = \begin{bmatrix}
        0.02 \\  0.02 \\ 0.02
    \end{bmatrix}.
\end{align*}

\begin{figure}
    \centering
    \begin{subfigure}{\columnwidth}
         \centering
         \includegraphics{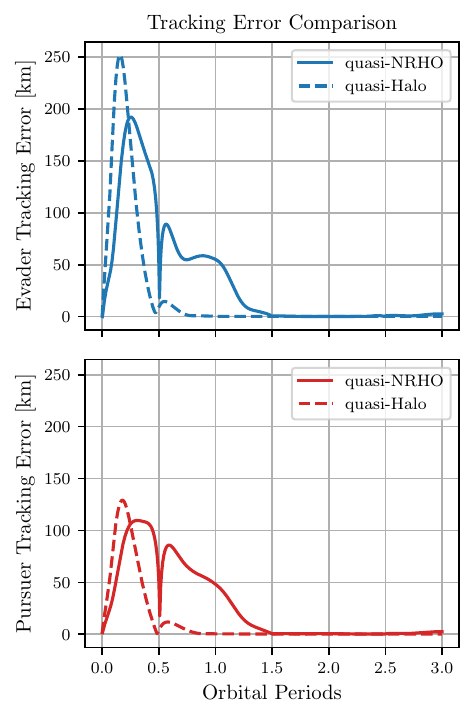}
         \caption{Evader and pursuer tracking errors over the quasi-NRHO and quasi-halo. The quasi-NRHO produces longer deviations, while the quasi-halo gonverges back to reference orbit quickier.}
         \label{fig:NRHO_halo_tracking}
     \end{subfigure}
     \begin{subfigure}{\columnwidth}
         \centering
        \includegraphics{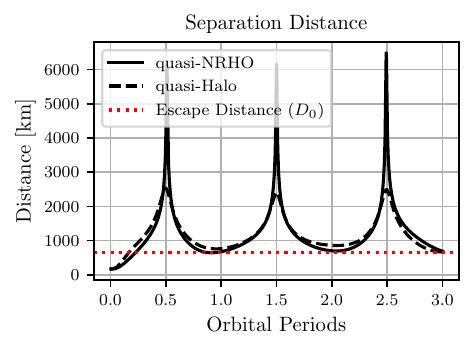}
         \caption{Separation distance over the quasi-NRHO and quasi-halo. The quasi-NRHO generates larger separation peaks because the close lunar passages amplify differences established before the encounter.}
         \label{fig:NRHO_halo_separation}
     \end{subfigure}
     \caption{Comparison between pursuit-evasion games on a quasi-NRHO and a quasi-halo. The close lunar passages of the quasi-NRHO produce sharper tracking deviations and amplify the relative state established before perilune, resulting in larger separation than in the quasi-halo case.}
\end{figure}

The tracking-error histories in Fig.~\ref{fig:NRHO_halo_tracking} show how strongly the trajectories depart from their respective phase-adjusted references. In the quasi-NRHO case, these deviations are closely associated with the close lunar passage, where relatively small changes in the incoming state can produce substantially different outgoing trajectories. By contrast, the quasi-halo case remains in a weaker gravitational environment, and its tracking deviations evolve less sharply.

As shown in Fig.~\ref{fig:NRHO_halo_separation}, the quasi-NRHO evader can exploit this sensitivity to generate a larger physical separation. However, the close lunar passage should not be interpreted primarily as the interval during which this separation is created. Rather, the lunar encounter amplifies differences in position and velocity that are established beforehand. The most consequential control actions must therefore be applied before the spacecraft enter the close-approach corridor. During this pre-encounter interval, the agents use thrust and phase adjustment to place themselves on favorable incoming trajectories. Once they enter the corridor, the short encounter timescale and strong state-transition sensitivity make late control corrections less effective, so the subsequent motion is largely determined by the incoming encounter geometry.

\begin{figure}
    \centering
    \includegraphics{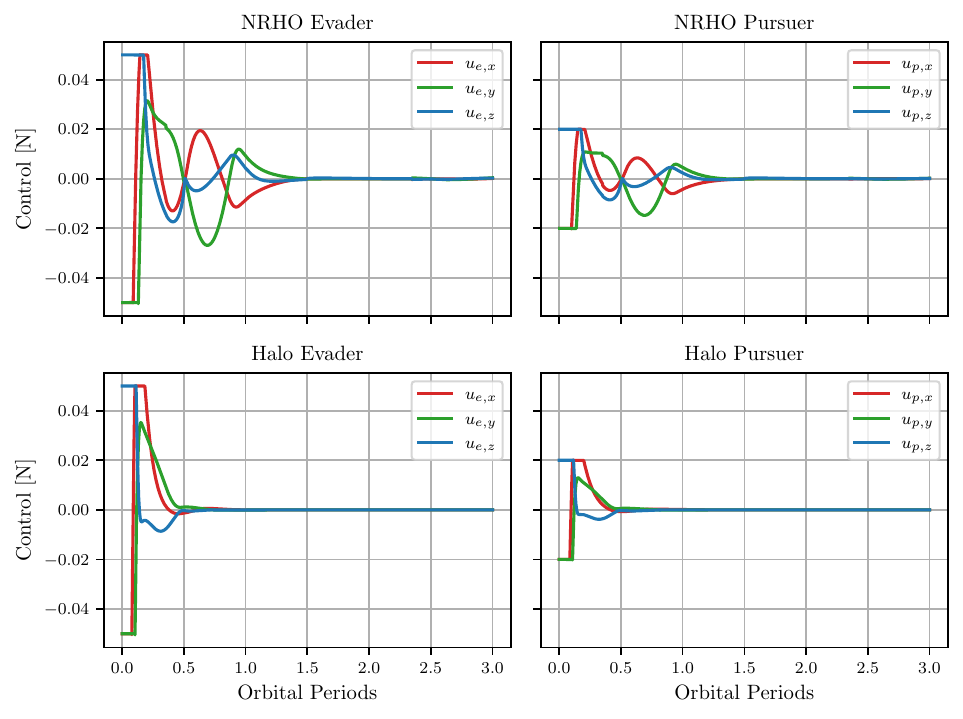}
    \caption{Control-effort comparison between the quasi-NRHO and quasi-halo cases. In the quasi-NRHO case, the main control actions occur before the close lunar passages, while the agents can still shape the incoming encounter geometry. The quasi-halo case shows a more shorter use of control because it does not experience the same short, strongly amplifying lunar encounters.}
    \label{fig:NRHO_halo_control}
\end{figure}

This temporal structure is reflected in the control-effort histories in Fig.~\ref{fig:NRHO_halo_control}. In the quasi-NRHO case, the relevant control effort is concentrated before the close lunar passage, while the agents can still modify the geometry of the encounter. Before periapsis, the evader tries to enter the lunar passage on a trajectory that leads to a favorable exit, while the pursuer tries to follow a similar outgoing path. Once the agents enter the close-approach corridor, they are left with limited maneuvering authority.

\begin{figure}
    \centering
    \includegraphics{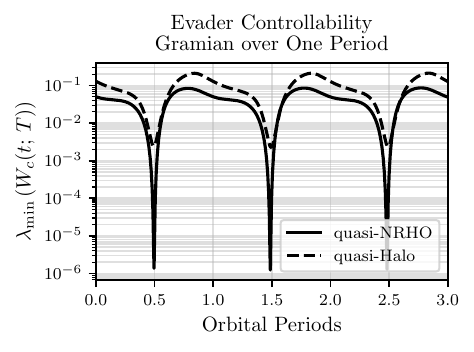}
    \caption{Comparison of the minimum controllability Gramian eigenvalue for the quasi-NRHO and quasi-halo. The sharp drops in the quasi-NRHO case show a stronger loss of control authority near the lunar encounters.}
    \label{fig:NRHO_halo_sensitivity}
\end{figure}

The sharp decrease in the controllability Gramian minimum eigenvalue near each lunar encounter in Fig.~\ref{fig:NRHO_halo_sensitivity} indicates that some state directions become weakly controllable. This loss of control authority is substantially stronger for the quasi-NRHO than for the quasi-halo, which helps explain why the encounter geometry established before perilune has a larger effect on the following separation. The close lunar passage can therefore create larger escape opportunities for the evader, but it also increases the risk that a pursuer entering with a favorable relative state will rapidly reduce the separation.

The quasi-halo remains farther from the Moon and therefore experiences weaker gravitational gradients. Its relative motion is shaped more continuously by thrust and two-dimensional phase adjustment rather than by a short, strongly amplifying lunar passage. Since the agents receive less assistance from the Moon dynamics, the maximum achievable separation is limited by their available control authority. Once the evader establishes sufficient separation, the pursuer cannot rely on a close lunar encounter to recover the lost distance. These two scenarios reveal a fundamental trade-off. The quasi-halo produces a more uniform pursuit-evasion interaction but more limited separation capability. The quasi-NRHO permits larger separation by exploiting the lunar dynamics, but its strategy must be established before the close approach and is more sensitive to the relative placement of the agents. In this case, the control selects the incoming geometry, the lunar passage amplifies it, and the resulting outgoing geometry determines the separation available over the subsequent orbital arc.

\section{Conclusion}
We investigated pursuit-evasion in cislunar space by formulating it as a zero-sum differential game under the CR3BP dynamics. The objective was to establish a control policy that allowed an evading spacecraft to remain close to its nominal orbit while maintaining a minimum safe separation from a pursuing spacecraft.

First, we introduced reference orbit phasing as an additional control variable, allowing each spacecraft to maneuver along its nominal trajectory without requiring large geometric departures from the reference orbit. For periodic orbits, phase control provided an additional along-track maneuverability, while for quasi-periodic orbits it extended naturally to two independent phase directions on the invariant torus. This additional degree of freedom enabled the evader to exploit the geometry of the reference manifold itself, creating separation while maintaining substantially smaller tracking errors than were possible using thrust alone.

In addition, we proposed a shared fictitious-time regularization for multi-agent systems. By extending Sundman regularization to a multi-agent setting, the method preserved synchronization between agents while automatically increasing numerical resolution during close lunar approaches. This removed inconsistencies associated with independently regularized spacecraft and enabled accurate discretization of joint trajectories over orbits with close encounters with the Moon.

The numerical experiments demonstrated that the geometry of the reference orbit played a fundamental role in defensive maneuvering. Quasi-periodic orbits provided greater maneuvering flexibility than periodic orbits by allowing motion over a two-dimensional invariant torus rather than a single curve. Moreover, comparisons between quasi-halo and quasi-NRHO families revealed a fundamental trade-off: trajectories that passed closer to the Moon generated larger escape opportunities by exploiting stronger nonlinear gravitational effects, but these same effects also increased the pursuer's ability to close the distance, resulting in greater encounter sensitivity. Orbit selection therefore became an integral component of defensive mission design.

Overall, these results extend pursuit-evasion theory from two-body orbital mechanics to the cislunar regime and provide a unified framework that combines nonlinear orbital dynamics, differential game theory, adaptive regularization, and reference manifold control. Future work will investigate imperfect state information, communication delays, multiple cooperating pursuers and evaders, and higher-fidelity dynamical models including perturbations beyond the CR3BP. Such extensions will further advance autonomous guidance and resilience for future missions operating in cislunar space.

\bibliography{sample}

\appendix
\section*{Appendix}

\subsection{Adapted Frame}\label{app:Frame}
In practice, the effectiveness of the reduction depends on the conditioning of the local frame
\begin{align}
    L = \left[\partial_\theta K\right(\theta)\quad f(K(\theta)) \quad W(\theta)].
\end{align}
This affects the numerical stability of the adapted-frame construction and of the subsequent correction step. Following Remark 3.3.1 in \cite{haro2021flow}, the adapted frame is not unique and may be modified by admissible transformations of its columns. In particular, we apply a constant linear transformation to the tangent block $\left[\partial_\theta K\right(\theta)\quad X_H(\theta)]$ and a constant scalar scaling to the bundle direction $W$. The goal is to normalize these directions on average over the torus, so that the columns of $L$ have comparable magnitude and improved conditioning. More precisely, let
\begin{align}
    \mathbf{T}_{i,j} = \left[\partial_\theta K_i\right(\theta_j)\quad f(K_i(\theta_j))]
\end{align}
where $\theta_j$ is the discretized torus parameterization angle and $i$ is the multi-shooting node index. We define the average Gram matrix of the tangent block by
\begin{align}
    M = \frac{1}{mN}\sum_{i=1}^m\sum_{j=1}^N \mathbf{T}_{i,j}^T\mathbf{T}_{i,j}.
\end{align}
Assuming $M$ is positive definite, let $C$ be its Cholesky factor, $M=C^TC$, and define $A=C^{-1}$, then $A^TMA=I$, and $\mathbf{T}_{i,j}A$ has a unit average Gram matrix over the discrete torus grid. Similarly, we define the average bundle norm 
\begin{align}
    M_W = \frac{1}{mN}\sum_{i=1}^m\sum_{j=1}^N ||W_{i,j}||^2_2,
\end{align}
where $W_{i,j}=W_i(\theta_j)$ and use the scaled bundle $Wb$, where $b=M_W^{-1/2}$, which has unit average squared norm. The final frame used in the computation is therefore
\begin{align}
    L_{i,j} = \left[\mathbf{T}_{i,j}A \quad W_{i,j}b\right]
\end{align}
This transformation does not change the subspaces spanned by the tangent and bundle directions. It only changes their coordinates inside the adapted frame. Therefore, it preserves the geometric content of the method while reducing the ill-conditioning of $L$ in practice.

\subsection{Kustaanheimo-Stiefel Regularized Dynamics} \label{app:KS}
We can regularize one of the two singularities in the CR3BP using the Kustaanheimo--Stiefel (KS) transformation \cite{oguri2024regularization}, an extension of the Levi-Civita planar regularization. In three dimensions, we must move to four dimensions to gain the extra mathematical dimension needed to smooth out the singularity.

To remove the singularity at the Moon, we first recenter the coordinate frame so the origin lies at the Moon center. Transforming the spatial coordinates alone is
insufficient; we must also regularize time. Because the spacecraft's velocity approaches infinity near the collision point, we introduce a fictitious time parameter $\tau$ via a Sundman transformation, $dt/d\tau = r_M$, which stretches the time step near the singularity and keeps the equations of motion well-behaved. We define $\mathbf{z} \in \mathbb{R}^4$ as the Cartesian position expressed in KS coordinates, and $\mathbf{y} = [\mathbf{z}, \mathbf{z}', H, t] \in \mathbb{R}^{10}$ as the extended state vector, where $\mathbf{z}'$ denotes the velocity with respect to the fictitious time $\tau$ and $H$ is the CR3BP Hamiltonian. The new translational equations of motion become
\begin{align}
    \mathbf{y}' = f_{KS}(\mathbf{y})+B_{KS}(\mathbf{y})\mathbf{u}
\end{align}
where
\begin{align}
    f_{KS}(\mathbf{y}) &:= \begin{bmatrix}
        \mathbf{z'}\\
        \frac{1}{2}h\mathbf{z}+L^TML\mathbf{z'}+\frac{1}{2}z^2L^T\mathbf{g}\\
        0\\
        z^2
    \end{bmatrix}, \quad B_{KS}(\mathbf{y}):=\frac{1}{m}\begin{bmatrix}
        0_{4\times3}\\
        \frac{1}{2}z^2L^TE_a\\
        2[\mathbf{z}']^TL^TE_a\\
        0_{1\times 3}
    \end{bmatrix}, \\
    E_a &= \begin{bmatrix}
        I_3\\
        0_{1\times 3}
    \end{bmatrix}, \quad h = H + \frac{1}{2}(x^2+y^2)+\frac{1-\mu}{r_E},\\
    M &= \begin{bmatrix}
        0 & 2 & 0 & 0\\
        -2 & 0 & 0 & 0\\
        0 & 0 & 0 & 0\\
        0 & 0 & 0 & 0
    \end{bmatrix}, \quad \mathbf{g} =\begin{bmatrix}
        x\\ y \\ 0 \\ 0
    \end{bmatrix} - \frac{1-\mu}{r_E^3}\begin{bmatrix}
        x+\mu\\ y\\ z\\ 0
    \end{bmatrix}, \quad L = \begin{bmatrix}
        z_1 & -z_2 & -z_3 & z_4\\
        z_2 & z_1 & -z_4 & -z_3\\
        z_3 & z_4 & z_1 & z_2\\
        z_4 & -z_3 & z_2 & -z_1
    \end{bmatrix}.
\end{align}

\subsection{Numerical Experiment Parameters}
Table~\ref{tab:simulation_parameters} summarizes the dynamical and numerical parameters used in the simulations. It includes the CR3BP normalization constants, the reference orbit initial conditions, and the main DDP settings.       
\begin{table}[h]
\centering
\caption{Dynamical and numerical parameters used in the simulations.}
\label{tab:simulation_parameters}
\renewcommand{\arraystretch}{1.1}
\begin{tabular}{@{}ll@{}}
\hline
Parameter & Value \\
\hline
CR3BP mass parameter, $\mu$
& $1.21506\times10^{-2}$ \\
Length unit, LU [km]
& $389703$ \\
Time unit, TU [s]
& $382981$ \\[0.2em]

NRHO initial state, $\mf X_0$ [LU]
& $\begin{bmatrix}
1.018659 & 0 & -0.179672 &
0 & -0.095814 & 0
\end{bmatrix}^\mr{T}$\\
NRHO period, T [TU]
& $1.466695$\\[0.2em]

Halo initial state, $\mf X_0$ [LU]
& $\begin{bmatrix}
1.088688 & 0 & -0.201828 &
0 & -0.206654 & 0
\end{bmatrix}^\mr{T}$ \\
Halo period, T [TU]
& $2.469518$ \\[0.2em]

Maximum DDP iterations
& $500$ \\
Number of discretization nodes
& $1000$ \\
Number of orbital periods
& $3$ \\
Number of replans
& $5$ \\
Convergence tolerance, $\varepsilon$
& $10^{-5}$ \\
Lower acceptance threshold, $\varepsilon_1$
& $0.5$ \\
Upper acceptance threshold, $\varepsilon_2$
& $2$\\
\hline
\end{tabular}
\end{table}

\end{document}